\documentclass[11pt,letterpaper]{article}

\usepackage[margin=1in]{geometry}
\usepackage{setspace}
\usepackage[T1]{fontenc}
\usepackage[utf8]{inputenc}
\usepackage{lmodern}
\usepackage{graphicx}
\usepackage{booktabs}
\usepackage{natbib}
\usepackage{authblk}
\usepackage{tikz}
\usepackage{enumerate}

\usepackage{amsmath,amssymb,amsthm,mathtools}
\IfFileExists{bbm.sty}{\usepackage{bbm}}{%
  \providecommand{\mathbbm}{\mathbf}}
\numberwithin{equation}{section}
\theoremstyle{plain}
\newtheorem{theorem}{Theorem}[section]
\newtheorem{lemma}[theorem]{Lemma}
\newtheorem{proposition}[theorem]{Proposition}
\newtheorem{corollary}[theorem]{Corollary}
\theoremstyle{definition}
\newtheorem{definition}{Definition}
\newtheorem{assumption}{Assumption}
\newtheorem{remark}{Remark}[section]

\IfFileExists{algorithm2e.sty}{%
  \usepackage[ruled,linesnumbered]{algorithm2e}}{}
\ifdefined\KwIn\else
\newcounter{algocf}
\newcommand{\AlgCaption}[1]{%
  \noindent\textbf{Algorithm \thealgocf: #1}\par\smallskip\hrule\medskip}
\newenvironment{algorithm}[1][t]{%
  \par\medskip\noindent\begin{minipage}{\linewidth}
  \refstepcounter{algocf}\let\caption\AlgCaption
  \hrule\medskip
  \def\;{\ifmmode\mskip\thickmuskip\else\par\fi}}{%
  \medskip\hrule\end{minipage}\par\medskip}
\newcommand{\KwIn}[1]{\noindent\textbf{Input: }#1\par}
\newcommand{\KwOut}[1]{\noindent\textbf{Output: }#1\par\smallskip}
\newcommand{\Return}[1]{\textbf{return }#1}
\newcommand{\AlgBody}[1]{%
  \begin{list}{}{\setlength{\leftmargin}{1.2em}%
    \setlength{\topsep}{0pt}\setlength{\parsep}{0pt}}\item[]#1\end{list}}
\newcommand{\If}[2]{\noindent\textbf{if }#1\textbf{ then}\par\AlgBody{#2}}
\newcommand{\For}[2]{\noindent\textbf{for }#1\textbf{ do}\par\AlgBody{#2}}

\fi
\usepackage{xcolor}
\usepackage[colorlinks,linkcolor=blue,citecolor=blue,
  urlcolor=blue,breaklinks]{hyperref}

\def\E{\mathbb{E}}
\def\P{\mathbb{P}}
\def\R{\mathbb{R}}
\def\N{\mathbb{N}}
\def\Xcal{\mathcal{X}}
\newcommand{\bx}{\mathbf{x}}
\newcommand{\bX}{\mathbf{X}}
\newcommand{\bY}{\mathbf{Y}}

\def\Hcal{\mathcal{H}}
\def\Ccal{\mathcal{C}}

\newcommand{\One}[1]{\mathbbm{1}\left\{#1\right\}}

\def\conch{\textnormal{\textsc{CONCH}}}

\DeclareMathOperator*{\argmax}{argmax}

\newcommand{\Var}{\operatorname{Var}}
\newcommand{\KL}{\operatorname{KL}}
\newcommand{\TV}{\operatorname{TV}}

\newcommand{\Op}{O_{\mathbb P}}
\newcommand{\splitconch}{\textnormal{\textsc{CONCH}-SPLIT}}
\newcommand{\detectconch}{\textnormal{\textsc{DTL-CONCH}}}
\newcommand{\clone}{\textnormal{\textsc{CLONE}}}

\usepackage[nameinlink,noabbrev]{cleveref}
\crefname{assumption}{assumption}{assumptions}

\newcommand{\papertitle}{Changepoint localization via nested exchangeability tests: from point estimates to confidence intervals}
\newcommand{\paperauthorA}{Rohan Hore}
\newcommand{\paperauthorB}{Aaditya Ramdas}
\newcommand{\affilOne}{Department of Statistics, Stanford University}

\newcommand{\corrEmail}{rhore@stanford.edu}

\title{\papertitle}
\author{\paperauthorA\thanks{Corresponding author: \corrEmail} }
\author{\paperauthorB}
\affil{\affilOne}
\date{\today}
\newcommand{\paperabstract}[1]{\begin{abstract}#1\end{abstract}}

\begin{document}
\maketitle
\paperabstract{
We study distribution-free uncertainty quantification for a changepoint estimate in an offline setting. We observe a sequence with a single changepoint and are given a preliminary estimate of its location. The pre- and post-change segments are independent and each consists of i.i.d.\ (or, more generally, exchangeable) observations from arbitrary and unknown distributions. We introduce Changepoint LOcalization via Nested Exchangeability testing (\clone{}), a novel method for constructing distribution-free confidence intervals around that estimate by testing exchangeability in nested windows centered at the reference estimate. Under these assumptions, \clone{} provides finite-sample coverage for any black-box reference estimator. Using a geometrically shrinking radius grid, \clone{} requires only $O(\log n)$ permutation tests. We characterize optimal scores for the framework and provide practical recommendations that result in sharp confidence intervals. Under mild assumptions, \clone{} confidence intervals have asymptotically vanishing normalized length. Simulations and experiments on image and text data support our theoretical findings.
}

\section{Introduction}
Suppose we observe a data sequence with a single changepoint. We study the problem of offline changepoint localization. A long line of classical statistical methods provides point estimates of the change location \citep{quandt1958estimation,kim1989likelihood}. More recently, modern machine learning methods, including random forests and neural networks, have also been used to estimate changepoint locations \citep{londschien2023random,li2024automatic}. However, these estimates typically do not come with a finite-sample reliability certificate. Our goal is to provide distribution-free uncertainty quantification for such estimates, turning a preliminary changepoint estimate into a confidence interval that contains the true changepoint with a user-specified coverage probability.

\subsection{Problem setup}
We now formalize the problem. Suppose we observe a sequence $\bX=(X_1,\ldots,X_n)\in\Xcal^n$ for some $n\ge2$, where $\Xcal$ is a general measurable space. We assume that the sequence $\bX$ admits a single changepoint, denoted by $\xi\in[n-1]$, where we write $[K]$ to denote $\{1,\ldots,K\}$ for any integer $K\in\N$. We make the following assumption on the joint distribution of $\bX$.

\begin{assumption}\label{ass:exchangeability}
We assume that the segments $(X_1,\ldots,X_\xi)$ and $(X_{\xi+1},\ldots,X_n)$ are independent and each exchangeable.
\end{assumption}

We call the changepoint genuine if the one-dimensional marginal distributions induced by the two segments are different. A canonical example is the i.i.d.\ changepoint model: the observations are drawn i.i.d.\ from $P_0$ before the change and i.i.d.\ from $P_1\ne P_0$ after the change. No further knowledge of the distributions, their separation, or any particular structure on $\Xcal$ is assumed. This in fact poses the main challenge in distribution-free changepoint localization.

Our goal goes beyond obtaining a point estimate of the changepoint $\xi$. We seek a collection of time points that contains $\xi$ with high probability. We formalize this goal below.

\begin{definition}\label{eq:DF-conf_defn}
For $\alpha\in(0,1)$, a measurable mapping $\Ccal_{n,1-\alpha}:\Xcal^n\to2^{[n-1]}$ is a distribution-free confidence set for a changepoint if
\[
\P_{\bX\sim P}(\xi\in\Ccal_{n,1-\alpha}(\bX))\ge1-\alpha,
\]
for every distribution $P$ with a changepoint at $\xi\in[n-1]$ and satisfying Assumption~\ref{ass:exchangeability}.
\end{definition}

\subsection{Existing work}
\label{sec:review}
Offline changepoint analysis has a long history; see \citet{truong2020selective,duggins2010parametric} for surveys. A significant part of this literature focuses only on detecting the change. Beginning with CUSUM \citep{page1955test}, a large literature has developed around modern detection procedures, including kernel changepoint procedures (KCP) \citep{arlot2019kernel}, ChangeForest \citep{londschien2023random}, and conformal martingale methods in the online setting \citep{vovk2003testing,volkhonskiy2017inductive}.

The goal of this paper is to localize the changepoint efficiently while quantifying uncertainty in its location. Likelihood-based approaches assume specific parametric families, such as Gaussian mean-shift or linear regression models, and derive point estimators and confidence intervals through likelihood theory \citep{kim1989likelihood,quandt1958estimation,gurevich2006guaranteed}.

Other approaches derive localization guarantees through asymptotic theory, including SMUCE \citep{frick2014multiscale} and regression-based methods \citep{xu2024change}. Bootstrap approaches \citep{cho2022bootstrap} quantify uncertainty in changepoint estimates but generally lack finite-sample validity and can be computationally intensive.

However, these classical approaches do not meet the requirements of modern applications. Today, data are complex and often scarce, and therefore may not satisfy the model assumptions or justify the large-sample approximations required by these approaches. This motivates the distribution-free formulation outlined in Definition~\ref{eq:DF-conf_defn}. Building on conformal $p$-values, recent works \citep{dandapanthula2025conformal,hore2026conformal} have developed such changepoint localization frameworks with provable finite-sample, distribution-free coverage guarantees. These approaches start from a score function to measure the plausibility of a change at each time point and construct a confidence set by testing, for each candidate location $t\in[n-1]$, whether $t$ is the true changepoint, retaining the locations that are not rejected.

These conformal frameworks do not immediately provide uncertainty quantification around an existing changepoint estimate. In predictive inference, by contrast, conformal methods are commonly viewed as calibration wrappers: they take a black-box prediction algorithm and quantify uncertainty around its predictions. Since the existing conformal localization approaches rely on $n-1$ separate tests, one for each candidate location, it is not immediately clear how to wrap them around existing point estimates.

\subsection{Preview of our approach: \clone{}}\label{sec:clone_preview}
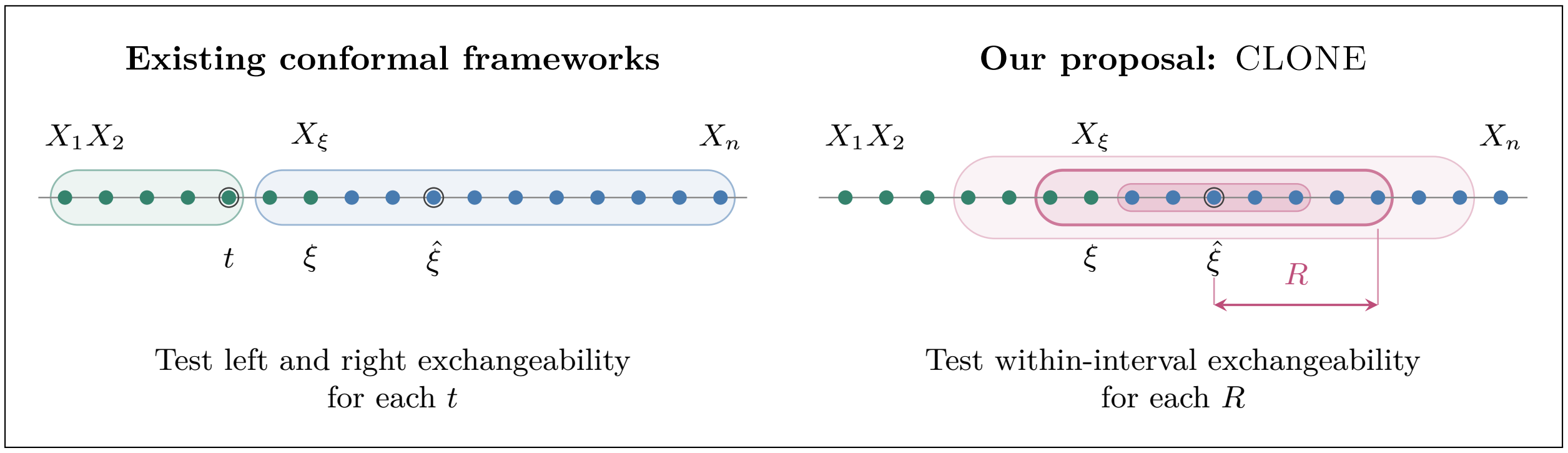
\begin{figure}[t]
    \centering
    \vspace{.4em}
    \begingroup
    \setlength{\fboxsep}{8pt}
    \definecolor{hypPre}{HTML}{35836D}
    \definecolor{hypPost}{HTML}{487BB0}
    \definecolor{hypPink}{HTML}{BD4D79}
    \fbox{\resizebox{\dimexpr\linewidth-2\fboxsep-2\fboxrule\relax}{!}{%
    \begin{tikzpicture}[x=.42cm,y=1cm,font=\fontsize{9}{11}\selectfont,
        timeline/.style={draw=black!45,line width=.45pt},
        heading/.style={font=\fontsize{10}{12}\selectfont\bfseries}]

    \newcommand{\HypTimeline}{%
        \draw[timeline] (-.65,0)--(16.65,0);
        \foreach \i in {0,...,16}{%
            \ifnum\i<7
                \fill[hypPre] (\i,0) circle (2.1pt);
            \else
                \fill[hypPost] (\i,0) circle (2.1pt);
            \fi
        }
        \foreach \x/\lab in {0/1,1/2,6/\xi,16/n}
            \node at (\x,.62) {$X_{\lab}$};
        \node at (6,-.62) {$\xi$};
        \node at (9,-.62) {$\hat\xi$};
        \draw[black!75,line width=.45pt] (9,0) circle (2.8pt);
    }

    \begin{scope}
        \node[heading] at (8,1.4) {Existing conformal frameworks};
        \draw[rounded corners=8pt,draw=hypPre!55,fill=hypPre!9,line width=.5pt]
            (-.35,-.28) rectangle (4.35,.28);
        \draw[rounded corners=8pt,draw=hypPost!55,fill=hypPost!9,line width=.5pt]
            (4.65,-.28) rectangle (16.35,.28);
        \HypTimeline
        \draw[black!75,line width=.45pt] (4,0) circle (2.8pt);
        \node at (4,-.62) {$t$};
        \node[align=center] at (8,-1.85)
            {Test left and right exchangeability\\for each $t$};
    \end{scope}

    \begin{scope}[xshift=8cm]
        \node[heading] at (8,1.4) {Our proposal: \clone{}};
        \draw[rounded corners=12pt,draw=hypPink!35,fill=hypPink!7,line width=.45pt]
            (2.65,-.42) rectangle (15.35,.42);
        \draw[rounded corners=8pt,draw=hypPink!75,fill=hypPink!16,line width=.85pt]
            (4.65,-.28) rectangle (13.35,.28);
        \draw[rounded corners=4pt,draw=hypPink!60,fill=hypPink!30,line width=.45pt]
            (6.65,-.14) rectangle (11.35,.14);
        \HypTimeline
        \draw[draw=hypPink!65,line width=.45pt] (9,-.82)--(9,-1.1);
        \draw[draw=hypPink!65,line width=.45pt] (13,-.32)--(13,-1.1);
        \draw[<->,>=stealth,draw=hypPink,line width=.65pt]
            (9,-1.1)--(13,-1.1)
            node[midway,above=2pt,text=hypPink] {$R$};
        \node[align=center] at (8,-1.85)
            {Test within-interval exchangeability\\for each $R$};
    \end{scope}
    \end{tikzpicture}%
    }}
    \endgroup
    \caption{A schematic comparison of the hypotheses tested: green and blue points indicate the true pre- and post-change segments, with changepoint $\xi$ and reference estimate $\hat\xi$. On the left, we have the hypotheses tested by \citet{hore2026conformal,dandapanthula2025conformal}. For each candidate $t$, the null asserts that the two disjoint segments determined by $t$ are independent and each exchangeable; the shaded regions mark these segments. On the right is our proposal: for each radius $R$, \clone{} tests whether the observations in the associated window around $\hat\xi$ are exchangeable. The pink regions illustrate nested windows, parametrized by radius $R$.}
    \label{fig:hypothesis-tests}
\end{figure}
A natural way to calibrate an existing changepoint estimate is to construct a confidence interval around it with distribution-free coverage for the true changepoint, thereby quantifying uncertainty in the estimate.

Suppose we are given a preliminary changepoint estimate $\hat\xi$. We consider nested intervals around it with decreasing radii, eventually reaching the singleton $\{\hat\xi\}$. Starting from the full range of candidate locations, we test exchangeability in successively smaller windows and continue shrinking whenever the test rejects. The intuition is simple: if the true changepoint $\xi$ lies outside a proposed interval, the associated window is homogeneous, so its exchangeability test should typically not reject. We return the preceding interval at the first non-rejection, or the singleton set if every test rejects. We call our proposed method Changepoint LOcalization via Nested Exchangeability tests (\clone{}). An illustration of the different hypothesis tests considered by existing conformal frameworks and our proposal is given in Figure~\ref{fig:hypothesis-tests}.

\begin{figure}[t]
    \centering
    \includegraphics[width=0.85\linewidth]{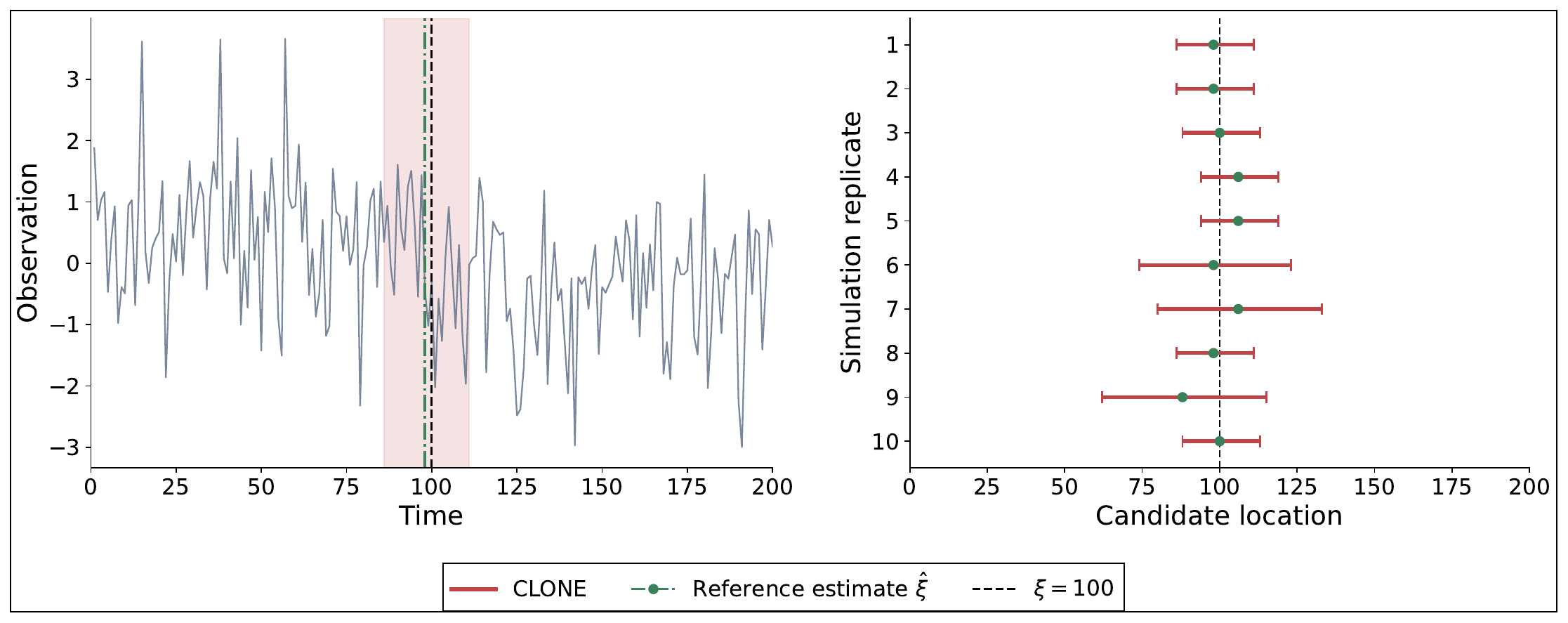}
    \caption{Gaussian mean shift localization with $n=200$, $\xi=100$, and pre- and post-change distributions $\mathcal N(0.5,1)$ and $\mathcal N(-0.5,1)$, respectively. Left: a simulated sequence with the true changepoint (dashed), the reference estimate (dash-dotted), and the $90\%$ \clone{} confidence interval (shaded). Right: \clone{} intervals across ten simulations, with green points marking the corresponding reference estimates. \clone{} quantifies uncertainty around each reference estimate through a contiguous confidence interval.}
    \label{fig:preview}
\end{figure}

The use of nested intervals around a point estimate has previously appeared in the conformal literature. In predictive inference, \citet{gupta2022nested} consider nested intervals around a point estimate and calibrate them efficiently to produce valid prediction intervals for the unseen test response. There, the nested conformal prediction intervals help measure the uncertainty around the initial estimate. Likewise, \clone{} provides uncertainty quantification for any given black-box changepoint estimator. A preview of the \clone{} framework is given in Figure~\ref{fig:preview}, where the confidence sets are contiguous intervals around the changepoint estimates and cover the true changepoint $\xi$ in most simulations.

\subsection{Our contributions}
The main contributions of this work are summarized below:
\begin{enumerate}[(i)]
\item We formally describe the \clone{} construction and establish its finite-sample validity. This coverage guarantee holds for any score used to test exchangeability within each window, providing the analyst with flexibility in designing their score. With a radius grid that shrinks geometrically, \clone{} requires only $O(\log n)$ permutation tests.

\item An efficient score results in tight confidence intervals. We characterize optimal scores for the individual exchangeability tests in \clone{} and give practical recommendations for approximating these scores.

\item We establish asymptotic sharpness of \clone{} confidence intervals for general scores in the i.i.d.\ changepoint model through an interpretable signal-to-noise ratio, without requiring consistent estimation of the underlying densities. When the pre- and post-change marginals are known, the \clone{} construction attains the rate given by the information-theoretic lower bound on cardinality.
\end{enumerate}

\paragraph{Organization of the paper.}
The rest of the paper is organized as follows. In Section~\ref{sec:clone}, we introduce \clone{} and establish its finite-sample validity. In Section~\ref{sec:efficiency}, we develop optimal and practical scores for the exchangeability tests and derive asymptotic localization rates. In Section~\ref{sec:simulations}, we present simulations and experiments on image and text data. In Section~\ref{sec:conch}, we review \conch{} and consider its interval hull as a baseline for comparison. We establish the same localization rate for the hull and show that, in the oracle setting, both methods attain the rate given by the information-theoretic cardinality lower bound. We conclude with a discussion in Section~\ref{sec:discussion}. All proofs are deferred to the appendices.
\section{Changepoint localization via nested exchangeability tests}\label{sec:clone}
In this section, we formally describe the \clone{} framework. Inspired by the rich literature on split conformal prediction \citep{papadopoulos2002inductive}, we adopt a split conformal approach to changepoint localization. We compute the reference estimate on one split and use the remaining observations for conformal calibration.

A random split need not place observations from both segments in each split. To preserve the temporal ordering and ensure that both splits contain a change, we use interlaced sample splitting.
We assume throughout this paper that the changepoint is away from the boundaries, i.e., $\xi\notin\{1,n-1\}$, so that both subsequences contain a change. We write $\mathcal{T}=\{2,\ldots,n-2\}$ to denote the set of possible changepoint locations.
Specifically, write $m=\lfloor n/2\rfloor$ and form the interlaced splits
\begin{equation}\label{eq:interlaced-splits}
 D_{n,1}=(X_{2i-1,n})_{i=1}^{\lceil n/2\rceil},
 \quad
 D_{n,2}=\bY_n=(Y_{1,n},\ldots,Y_{m,n}),
 \quad Y_{i,n}=X_{2i,n}.
\end{equation}
 Their changepoint indices are $\xi''=\lceil\xi/2\rceil$ and $\xi'=\lfloor\xi/2\rfloor$, respectively, with $|\xi''-\xi'|\le1$. We refer to $D_{n,1}$ and $D_{n,2}$ as the training and calibration splits, respectively.

First, we use $D_{n,1}$ to learn a reference estimate $\hat\xi_{n/2}\in[m-1]$. We write $\hat\xi_{n/2}$ to explicitly encode the dependence of the estimate on approximately half of the data.

Next, we calibrate $\hat\xi_{n/2}$ on the calibration split by considering nested intervals around this estimate. For an integer radius $R\ge0$, define
\begin{equation}\label{eq:clone-window}
 I_R=\{t\in[m-1]:|t-\hat\xi_{n/2}|\le R\}.
\end{equation}
If $\xi'\notin I_R$, then the observations $(Y_{L_R,n},\ldots,Y_{U_R,n})$ are exchangeable conditional on $D_{n,1}$, where $L_R=\max\{1,\hat\xi_{n/2}-R\}$ and $U_R=\min\{m,\hat\xi_{n/2}+R+1\}$. This window contains $k_R=U_R-L_R+1\le2R+2$ observations and motivates hypotheses of the form
\[
 \Hcal_{0,I_R}:
 (Y_{L_R,n},\ldots,Y_{U_R,n})
 \text{ is exchangeable conditional on }D_{n,1}.
\]
Consider a strictly decreasing integer sequence $R_0>R_1>\cdots>R_J=0$, which may be chosen based on the training split $D_{n,1}$, where $R_0=\max\{\hat\xi_{n/2},m-1-\hat\xi_{n/2}\}$. By construction, $I_{R_0}=[m-1]$ always contains $\xi'$. Moreover, the hypotheses $\Hcal_{0,I_R}$ are nested in $R$: if $R_1<R_2$ and $\Hcal_{0,I_{R_1}}$ is false, then $\Hcal_{0,I_{R_2}}$ must also be false. We therefore test the windows in decreasing order of radius, stopping at the first non-rejection and retaining the preceding interval.

To test $\Hcal_{0,I_R}$, we first define a window-specific score $S_R:\Xcal^{k_R}\to\R$, which may depend on the training split, with larger values providing greater evidence against exchangeability. Define the conformal $p$-value
\begin{equation}\label{eq:window-pvalue}
 p_{\mathrm{window},R}
 =\frac1{|\mathcal S_{k_R}|}
 \sum_{\pi\in\mathcal S_{k_R}}
 \One{S_R(\pi(Y_{L_R,n},\ldots,Y_{U_R,n}))
 \ge S_R(Y_{L_R,n},\ldots,Y_{U_R,n})},
\end{equation}
where we write $\mathcal S_{k_R}$ to denote the permutation group on $k_R$ items. Define the stopping rule
\[
 \hat j_\alpha=\min\left(
 \{j\in[J]:p_{\mathrm{window},R_j}>\alpha\}
 \cup\{J+1\}\right).
\]
We return the calibration confidence interval $I_{R_{\hat j_\alpha-1}}$. If $p_{\mathrm{window},R_j}\le\alpha$ for every $j\in[J]$, then $\hat j_\alpha=J+1$, resulting in the singleton calibration interval $I_0=\{\hat\xi_{n/2}\}$.

This gives a confidence interval on the calibration subsequence. To map back to the original timeline, let $B_t=\{2t,2t+1\}$ for $t\in[m-1]$ and observe that $\xi\in B_{\xi'}$ by construction. We report the resulting confidence interval
\begin{equation}\label{eq:clone-map}
 I^{\clone}_{n,1-\alpha}
 =\bigcup_{t\in I_{R_{\hat j_\alpha-1}}}B_t.
\end{equation}
The procedure is summarized in Algorithm~\ref{alg:clone}.

\begin{algorithm}[t]
\caption{\clone{}: changepoint localization via nested exchangeability tests}\label{alg:clone}
\KwIn{Calibration sequence $\bY_n$; training-based reference
$\hat\xi_{n/2}$, radius sequence $R_0>\cdots>R_J=0$, and window
scores $S_R$; level $\alpha$.}
\KwOut{A confidence interval on the original timeline.}
$\hat j_\alpha\leftarrow J+1$\;
\For{$j=1,\ldots,J$}{
  Compute $p_{\mathrm{window},R_j}$ using
  \eqref{eq:window-pvalue}\;
  \If{$p_{\mathrm{window},R_j}>\alpha$}{
    $\hat j_\alpha\leftarrow j$\;
    \textbf{break}\;
  }
}
$I^{\clone}_{n,1-\alpha}
 \leftarrow\bigcup_{t\in I_{R_{\hat j_\alpha-1}}}B_t$\;
\Return{$I^{\clone}_{n,1-\alpha}$}\;
\end{algorithm}

\begin{theorem}[Finite-sample coverage]
\label{thm:clone-valid}
Fix $\alpha\in(0,1)$ and $\xi\in\mathcal{T}$. The \clone{} confidence interval defined in Algorithm~\ref{alg:clone} satisfies
\begin{equation}\label{eq:clone-valid}
 \P_{\bX\sim P}(\xi\in I^{\clone}_{n,1-\alpha})\ge1-\alpha,
\end{equation}
for every distribution $P$ with a changepoint at $\xi$ and satisfying Assumption~\ref{ass:exchangeability}.
\end{theorem}

For large windows, enumerating the permutation group $\mathcal S_{k_R}$ to compute~\eqref{eq:window-pvalue} can be impractical. However, one need not enumerate all permutations in $\mathcal S_{k_R}$ to compute a valid conformal $p$-value.  Instead, one may use Monte Carlo calibration: for each tested window, draw fresh $\pi_1,\ldots,\pi_B\overset{\mathrm{iid}}{\sim}\mathrm{Unif}(\mathcal S_{k_R})$, independently of the observed data $\bX$, and compute
\begin{equation}\label{eq:window-mc-pvalue}
 \hat p_{\mathrm{window},R}
 =\frac1{B+1}\left(
 1+\sum_{k=1}^{B}\One{
 S_R(\pi_k(Y_{L_R,n},\ldots,Y_{U_R,n}))
 \ge S_R(Y_{L_R,n},\ldots,Y_{U_R,n})
 }
 \right).
\end{equation}
The finite-sample coverage guarantee continues to hold with these Monte Carlo $p$-values.

\begin{corollary}
\label{cor:clone-mc-valid}
Under the assumptions of Theorem~\ref{thm:clone-valid}, fix $B\ge1$ and replace $p_{\mathrm{window},R_j}$ in Algorithm~\ref{alg:clone} by $\hat p_{\mathrm{window},R_j}$ from \eqref{eq:window-mc-pvalue}. The returned confidence interval satisfies \eqref{eq:clone-valid}, where probability is taken over both the data $\bX$ and the sampled permutations.
\end{corollary}

The validity guarantees of \clone{} intervals hold for any window-specific score functions, offering great flexibility in score design. The choice of score still affects the width of the \clone{} interval. Before studying this in detail, we make a few remarks.

\begin{remark}[Conformal $p$-values with exact validity]
The full-permutation $p$-value in~\eqref{eq:window-pvalue} is only guaranteed to be super-uniform. By adding boundary randomization, we may define
\begin{align}\label{eq:window-pvalue-exact}
 \bar p_{\mathrm{window},R}
 =\frac1{|\mathcal S_{k_R}|}\biggl(
 &\sum_{\pi\in\mathcal S_{k_R}}
 \One{S_R(\pi(Y_{L_R,n},\ldots,Y_{U_R,n}))>
 S_R(Y_{L_R,n},\ldots,Y_{U_R,n})}\nonumber\\
 &+V_R\sum_{\pi\in\mathcal S_{k_R}}
 \One{S_R(\pi(Y_{L_R,n},\ldots,Y_{U_R,n}))=
 S_R(Y_{L_R,n},\ldots,Y_{U_R,n})}\biggr),
\end{align}
where $V_R\sim\mathrm{Unif}(0,1)$ is independent of the data and permutations. Under the null $\Hcal_{0,I_R}$, this $p$-value is exactly uniform conditional on the training split $D_{n,1}$ and the unordered window.
\end{remark}

\begin{remark}[Geometrically shrinking radius grid]\label{rem:geometric}
Since the hypotheses are nested, \clone{} can be implemented using a simple geometrically shrinking radius grid: starting from $R_0$, we recursively define $R_{j+1}=\lfloor R_j/2\rfloor$ until zero is reached. In this case, Algorithm~\ref{alg:clone} uses at most $1+\lfloor\log_2R_0\rfloor=O(\log n)$ permutation tests. In contrast, the existing conformal frameworks test $\Hcal_{0,t}$ for each $t\in[n-1]$, resulting in $n-1$ permutation tests. Note that this grid choice affects the resolution at which we perform the changepoint localization and may affect the resulting sharpness of the confidence interval; the coverage guarantee is unaffected.
\end{remark}
\section{Designing scores for \clone{}}\label{sec:efficiency}
An efficient window-specific score can lead to more powerful tests of exchangeability and thus narrower confidence intervals. In this section, we derive an expression for the optimal score and give practical approximations.

\subsection{Optimal window-specific scores}
To obtain an expression for an optimal score, we specialize to the i.i.d.\ changepoint model. Suppose that the pre-change observations $X_1,\ldots,X_\xi$ are drawn i.i.d.\ from a density $f_0$ and the post-change observations $X_{\xi+1},\ldots,X_n$ are drawn i.i.d.\ from a density $f_1$, with $f_0\neq f_1$. We assume that the corresponding distributions are mutually absolutely continuous.

We write $\Hcal_{0,\xi}$ to denote the hypothesis that Assumption~\ref{ass:exchangeability} holds with a changepoint at $\xi$. The corresponding calibration changepoint is at $\xi'$. We write $\P_\xi$ and $\E_\xi$ for probability and expectation under a joint distribution satisfying Assumption~\ref{ass:exchangeability} with a changepoint at $\xi$.

To obtain a narrow confidence interval, a powerful window-specific score should reject $\Hcal_{0,I_R}$ whenever the calibration changepoint $\xi'$ belongs to $I_R$. Write $j_\star=\max\{j\in\{0,\ldots,J\}:\xi'\in I_{R_j}\}$ for the largest index whose interval contains the calibration changepoint. One natural strategy for finding an optimal score is to consider the following objective:
\[
 \E_\xi\left[
 \sum_{j=1}^{j_\star}
 \One{p_{\mathrm{window},R_j}(S_{R_j})\le\alpha}
 \,\middle|\,D_{n,1}
 \right]=\sum_{j=1}^{j_\star}
 \P_\xi\left(
 p_{\mathrm{window},R_j}(S_{R_j})\le\alpha
 \,\middle|\,D_{n,1}
 \right),
\]
where $p_{\mathrm{window},R}(S_R)$ denotes the conformal $p$-value in~\eqref{eq:window-pvalue} computed with the window-specific score $S_R$. This objective can be interpreted as the expected number of correct rejections over the $j_\star$ nested hypotheses and serves as a surrogate for the final interval width. 

\begin{theorem}
\label{thm:window-np}
Consider the setting and notation defined above. Fix an integer $R\ge0$ and suppose $\xi'\in I_R$. Conditional on $D_{n,1}$ and the unordered observations $\{Y_{L_R,n},\ldots,Y_{U_R,n}\}$, a most powerful level-$\alpha$ test of $\Hcal_{0,I_R}$ against the alternative $\Hcal_{0,\xi}$ rejects for large values of
\begin{equation}\label{eq:window-oracle-score}
 S^\star_R(Y_{L_R,n},\ldots,Y_{U_R,n})
 =\sum_{i=L_R}^{\xi'}\ell(Y_{i,n}),
 \qquad \text{where }\ell=\log(f_0/f_1),
\end{equation}
where an empty sum equals zero. Consequently, using the conformal $p$-value~\eqref{eq:window-pvalue-exact} to attain the Type~I error level $\alpha$ exactly, we have, for any collection of window-specific scores $(S_{R_j})_j$,
\[
 \E_\xi\left[
 \sum_{j=1}^{j_\star}
 \One{\bar p_{\mathrm{window},R_j}(S_{R_j})\le\alpha}
 \,\middle|\,D_{n,1}
 \right]\le\E_\xi\left[
 \sum_{j=1}^{j_\star}
 \One{\bar p_{\mathrm{window},R_j}(S^\star_{R_j})\le\alpha}
 \,\middle|\,D_{n,1}
 \right].
\]
\end{theorem}
Therefore, $S^\star_R$ is the optimal score for the individual exchangeability test underlying the \clone{} framework, but it requires knowledge of both the log-likelihood ratio $\ell$ and the calibration changepoint $\xi'$, which are unknown in practice. Nevertheless, it motivates concrete and practical window-specific score constructions. We discuss these choices next.

\subsection{Practical window-specific scores}\label{sec:practical-score}
In practice, we first replace $\xi'$ by the reference estimate $\hat\xi_{n/2}$ computed on the training split $D_{n,1}$:
\[
 \hat\xi_{n/2}\in\argmax_{s\in[m-1]}\sum_{i=1}^s\ell(X_{2i-1,n}).
\]
Recalling that $\ell=\log(f_0/f_1)$, one can observe that $\hat\xi_{n/2}$ is essentially a maximum likelihood estimate (MLE) of the changepoint on the training split. We write $\hat\xi_{n/2}$ to explicitly encode the dependence of the estimate on approximately half of the data. This gives the score $\sum_{i=L_R}^{\hat\xi_{n/2}}\ell(Y_{i,n})$, which can be used when the oracle log-likelihood ratio $\ell$ is known. We refer to this score as the oracle LLR window score in this and later sections.

By construction, we always have $L_R\le\hat\xi_{n/2}<U_R$. Since the total window sum is unchanged by permutation, the oracle score $\sum_{i=L_R}^{\hat\xi_{n/2}}\ell(Y_{i,n})$ and the score
\[
 \frac{\sum_{i=L_R}^{\hat\xi_{n/2}}\ell(Y_{i,n})}{\hat\xi_{n/2}-L_R+1}
 -
 \frac{\sum_{i=\hat\xi_{n/2}+1}^{U_R}\ell(Y_{i,n})}{U_R-\hat\xi_{n/2}}
\]
produce the same permutation $p$-values. The latter has a natural interpretation: it is the difference in mean log-likelihood ratio between the two segments in the current window, treating $\hat\xi_{n/2}$ as the changepoint. Both denominators are positive for every radius $R\ge0$, and larger values indicate greater evidence of a change at the reference estimate. Henceforth, we use this mean-difference form throughout the remaining sections.

In practice, we will also not know $\ell$. Therefore, we may replace it by an estimate $\hat\ell_n$, which motivates the learned LLR window score
\begin{equation}\label{eq:window-statistic}
 S_R(Y_{L_R,n},\ldots,Y_{U_R,n})
 =
 \frac{\sum_{i=L_R}^{\hat\xi_{n/2}}\hat\ell_n(Y_{i,n})}
      {\hat\xi_{n/2}-L_R+1}
 -
 \frac{\sum_{i=\hat\xi_{n/2}+1}^{U_R}\hat\ell_n(Y_{i,n})}
      {U_R-\hat\xi_{n/2}}.
\end{equation}
The reference estimate is correspondingly obtained by maximizing the learned-score partial sum $\sum_{i=1}^s\hat\ell_n(X_{2i-1,n})$ over $s\in[m-1]$.
To obtain $\hat\ell_n$, one may first estimate the pre- and post-change densities $\hat f_{0,n}$ and $\hat f_{1,n}$ and take their log-likelihood ratio, using the training data alone or separate independent labeled data whenever available. Alternatively, one can learn $\hat\ell_n$ directly by training a classifier that distinguishes between the estimated pre- and post-change segments and then using its logits as an estimate. Both the learned function $\hat\ell_n$ and the reference estimate are held fixed during the conformal calibration on $D_{n,2}$.

\subsection{Asymptotic sharpness of \clone{} intervals}\label{sec:clone-sharpness}
We now study the sharpness of \clone{} confidence sets in an asymptotic setting. Specifically, consider a generalization of the i.i.d.\ changepoint model, with independent observations satisfying
\begin{equation}\label{eq:iid-model}
 X_{1,n},\ldots,X_{\xi_n,n}
 \overset{\mathrm{iid}}{\sim}P_{0,n},
 \qquad
 X_{\xi_n+1,n},\ldots,X_{n,n}
 \overset{\mathrm{iid}}{\sim}P_{1,n},
\end{equation}
where the pre- and post-change distributions may vary with the sample size. This allows local alternatives, under which $P_{0,n}$ and $P_{1,n}$ become increasingly similar as $n\to\infty$. We write $f_{0,n}$ and $f_{1,n}$ for their densities with respect to a common dominating measure, and $\ell_n=\log(f_{0,n}/f_{1,n})$ for the oracle log-likelihood ratio (LLR). As noted above, we do not require knowledge of $\ell_n$ to compute the \clone{} confidence interval and instead use an estimate $\hat\ell_n$ in its place.

As before, we adopt interlaced sample splitting, dividing the sequence $\bX_n=(X_{1,n},\ldots,X_{n,n})$ into the training split $D_{n,1}$ and the calibration split $D_{n,2}$. Their changepoints are located at $\xi_n''=\lceil\xi_n/2\rceil$ and $\xi_n'=\lfloor\xi_n/2\rfloor$, respectively. For the asymptotic analysis in this section, we learn $\hat\ell_n:\Xcal\to\R$ from a separate dataset $D_n'$ that is independent of $(D_{n,1},D_{n,2})$. Any randomness used in learning $\hat\ell_n$ is included in $D_n'$. The asymptotic results below use the full-permutation $p$-value in~\eqref{eq:window-pvalue}.

Using the training split $D_{n,1}$, we first define the reference estimate
\begin{equation}\label{eq:MLE}
 \hat\xi_{n/2}\in\argmax_{s\in[m-1]}
       \sum_{i=1}^{s}\hat\ell_n(X_{2i-1,n}),
\end{equation}
with a fixed measurable rule for resolving ties. Conditional on $(D_n',D_{n,1})$, both $\hat\ell_n$ and $\hat\xi_{n/2}$ are fixed. We then compute the conformal $p$-values on $D_{n,2}$ using the window-specific score $S_R$ defined in~\eqref{eq:window-statistic} and construct the \clone{} confidence interval.

To state the result, we introduce some notation.
For $j\in\{0,1\}$, define the unconditional score moments $\mu_{j,n}=\E_{D_n',\,X\sim P_{j,n}}[\hat\ell_n(X)]$ and $\sigma_{j,n}^2=\Var_{D_n',\,X\sim P_{j,n}}(\hat\ell_n(X))$, where $X$ is drawn independently of $D_n'$. These moments account for both the randomness of the learned function and variation within the pre- or post-change segments. To establish the sharpness guarantees, we impose the following conditions.

\begin{assumption}
\label{ass:learned-score}
We assume that
\begin{enumerate}[(i)]
 \item The changepoint location sequence $(\xi_n)_n$ satisfies $\xi_n/n\to\tau\in(0,1)$.

 \item There exists $\kappa\in(0,1/2]$ such that, for all sufficiently large $n$, $\mu_{0,n}>0>\mu_{1,n}$ and
 \begin{equation}\label{eq:centering}
  \mu_{0,n}\wedge(-\mu_{1,n})
  \ge\kappa(\mu_{0,n}-\mu_{1,n}).
 \end{equation}
 Further, $\sigma_{0,n}^2,\sigma_{1,n}^2<\infty$
 for every $n$.

 \item $\hat\ell_n$ satisfies the following stability condition: as $n\to\infty$,
 \begin{equation}\label{eq:stability}
  \varepsilon_n:=
  \frac{
   \Var_{D_n'}\!\left(
    \E_{X\sim P_{0,n}}[\hat\ell_n(X)\mid D_n']
   \right)+\Var_{D_n'}\!\left(
    \E_{X\sim P_{1,n}}[\hat\ell_n(X)\mid D_n']
   \right)}
  {(\mu_{0,n}-\mu_{1,n})^2}
  \longrightarrow0.
 \end{equation}
\end{enumerate}
\end{assumption}

Condition (i) is a mild assumption on the changepoint location. It ensures that the changepoint stays away from the boundaries of the timeline, so that both pre- and post-change segments cover a non-vanishing fraction of the sequence.

Condition (ii) is an assumption on the quality of the score $\hat\ell_n$. While it may not accurately estimate the oracle LLR, it should distinguish the two segments in expectation: pre-change observations receive a positive mean score, while post-change observations receive a negative mean score. The gap $\mu_{0,n}-\mu_{1,n}$ measures the separation of the pre- and post-change distributions through $\hat\ell_n$, and the condition~\eqref{eq:centering} requires each mean to be separated from zero by at least a fixed fraction of this gap. 

Finally, condition (iii) controls the randomness introduced by $\hat\ell_n$. It requires the conditional pre- and post-change score means to have small variances relative to the square of the unconditional mean separation gap from condition (ii). If $\hat\ell_n$ is deterministic, as with the oracle LLR, this condition is automatically satisfied.

\begin{remark}
\label{rem:centering}
For a fixed reference index, adding a deterministic constant to $\hat\ell_n$ leaves the \clone{} $p$-values unchanged. Subtracting the population midpoint $(\mu_{0,n}+\mu_{1,n})/2$ satisfies~\eqref{eq:centering} with $\kappa=1/2$. However, this centering may change the reference estimate in~\eqref{eq:MLE} and the ``maximum likelihood'' interpretation attached to it.
\end{remark}

Next, we define the signal-to-noise ratio (SNR) as
\begin{equation}\label{eq:SNR}
 \mathrm{SNR}_n
 =\frac{\mu_{0,n}-\mu_{1,n}}
 {(\sigma_{0,n}^2+\sigma_{1,n}^2)^{1/2}},
\end{equation}
with $\mathrm{SNR}_n=+\infty$ when the denominator is zero. The corresponding localization scale appearing in the results below is $r_n=1\vee\mathrm{SNR}_n^{-2}$. A weaker mean separation or greater score variability increases $r_n$, as one would naturally expect. Under local alternatives, this scale may diverge. The following proposition shows that the reference estimate localizes the changepoint at this scale.

\begin{proposition}
\label{prop:MLE}
Under the model~\eqref{eq:iid-model} and
Assumption~\ref{ass:learned-score}, for all sufficiently
large $n$ and every integer $u\ge2$,
\begin{equation}\label{eq:MLE-tail}
 \P(|\hat\xi_{n/2}-\xi_n'|>u)
 \le
 \frac{4\varepsilon_n}{\kappa^2}+
 \frac{4}{\kappa^2(u-1)}
 \frac{\sigma_{0,n}^2+\sigma_{1,n}^2}
      {(\mu_{0,n}-\mu_{1,n})^2}.
\end{equation}
Consequently, $|\hat\xi_{n/2}-\xi_n'|=\Op(r_n)$ and
$|2\hat\xi_{n/2}-\xi_n|=\Op(r_n)$.
\end{proposition}
The next theorem establishes the same localization rate for the \clone{} interval. As stated earlier in Remark~\ref{rem:geometric}, we use the geometric radius sequence $R_{j+1}=\lfloor R_j/2\rfloor$, starting from $R_0=\max\{\hat\xi_{n/2},m-1-\hat\xi_{n/2}\}$ and continuing until zero is reached.

\begin{theorem}
\label{thm:clone-sharp}
Under the model~\eqref{eq:iid-model} and the construction defined above, suppose Assumption~\ref{ass:learned-score} holds, with $r_n/n\to0$. Then, for every fixed $\alpha\in(0,1)$,
\begin{equation}\label{eq:clone-rate}
 |I^{\clone}_{n,1-\alpha}|=\Op(r_n),
 \qquad
 \frac{|I^{\clone}_{n,1-\alpha}|}{n-1}
 \xrightarrow{\P}0.
\end{equation}
\end{theorem}
Thus, under the stated assumptions, $\mathrm{SNR}_n\gg n^{-1/2}$ ensures that the \clone{} intervals have asymptotically vanishing normalized length.
\section{Experiments}\label{sec:simulations}
We now evaluate \clone{} in simulated and real data experiments. Throughout, we use interlaced sample splitting, a target confidence level of $90\%$, and a radius grid that halves at each shrinking step. Each Monte Carlo $p$-value uses $1000$ sampled permutations. 

\subsection{Gaussian mean shift localization}\label{sec:clone-gaussian-simulation}
We start by evaluating \clone{} in the Gaussian mean shift setting. We observe $n=500$ independent samples with a changepoint at $\xi=250$. The pre- and post-change distributions are $\mathcal N(\delta,1)$ and $\mathcal N(-\delta,1)$, respectively, with $\delta\in\{0.1,0.2,0.3,0.4,0.6,0.8,1\}$. Smaller $\delta$ corresponds to a weaker change, which is harder to localize. As $\delta$ increases, we thus expect narrower confidence intervals. We estimate the reference changepoint on the odd split by maximum likelihood and use the estimated segments to learn the densities. We then use the learned LLR window score for conformal calibration on the even split. We simulate $200$ independent sequences at each $\delta$.

\begin{figure}[t]
    \centering
    \includegraphics[width=\linewidth]{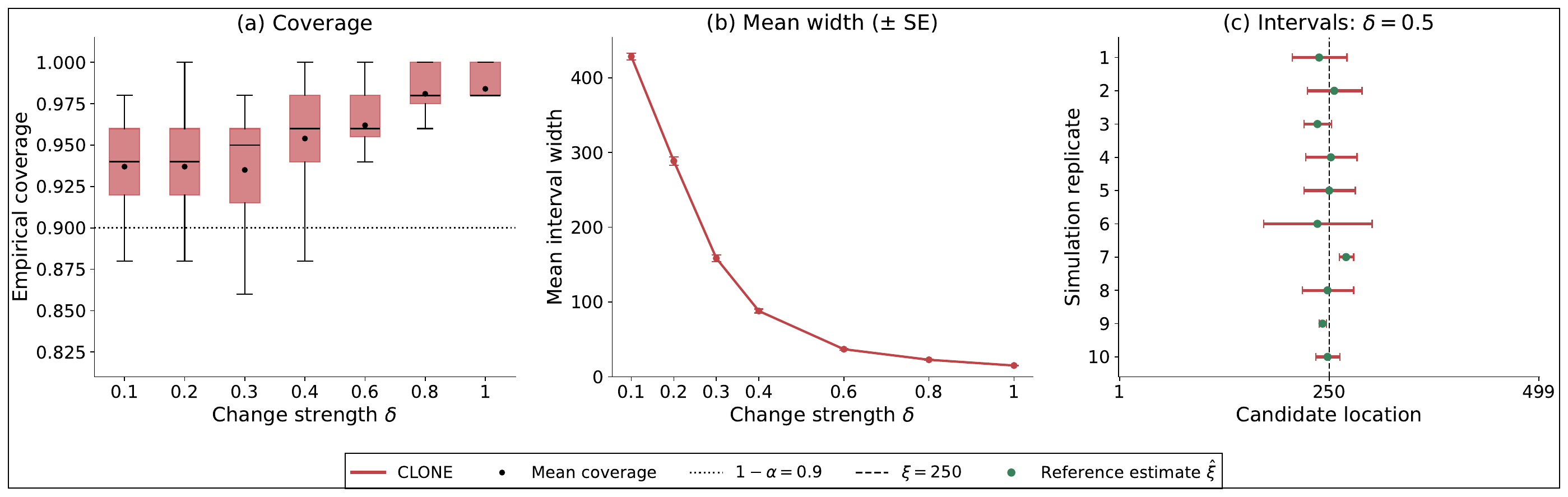}
    \caption{Gaussian mean shift: \clone{} maintains the nominal coverage and returns narrower intervals as the change strength increases. The intervals quantify uncertainty around the reference estimates, shown in green.}
    \label{fig:clone-gaussian-delta}
\end{figure}

Our findings are summarized in Figure~\ref{fig:clone-gaussian-delta}. In the left panel, we show boxplots of empirical coverage at each $\delta$. Average empirical coverage is at least $90\%$ in these simulations, in line with the theoretical guarantee. In the middle panel, we plot the average interval width with Monte Carlo standard errors. As $\delta$ increases, \clone{} returns narrower intervals. In the rightmost panel, we fix $\delta=0.5$ and plot intervals from ten simulated sequences, with green points marking the corresponding reference estimates.

\subsection{DomainNet: localizing domain shift}\label{sec:domainnet_expt}
Next, we consider domain shift localization using the DomainNet dataset \citep{peng2019moment}. We simulate sequences of $n=800$ images, with images from the real domain before $\xi=350$ and the sketch domain afterward. All images are converted to grayscale to remove color cues and make localization more challenging.

We train a CNN with four blocks on an independent labeled sample and use the classifier logits to construct the learned LLR score. We estimate the reference changepoint from the odd-indexed observations and calibrate \clone{} on the even split. We simulate ten sequences and compute the \clone{} intervals at the $90\%$ confidence level.

\begin{figure}[t]
    \centering
    \includegraphics[width=\linewidth]{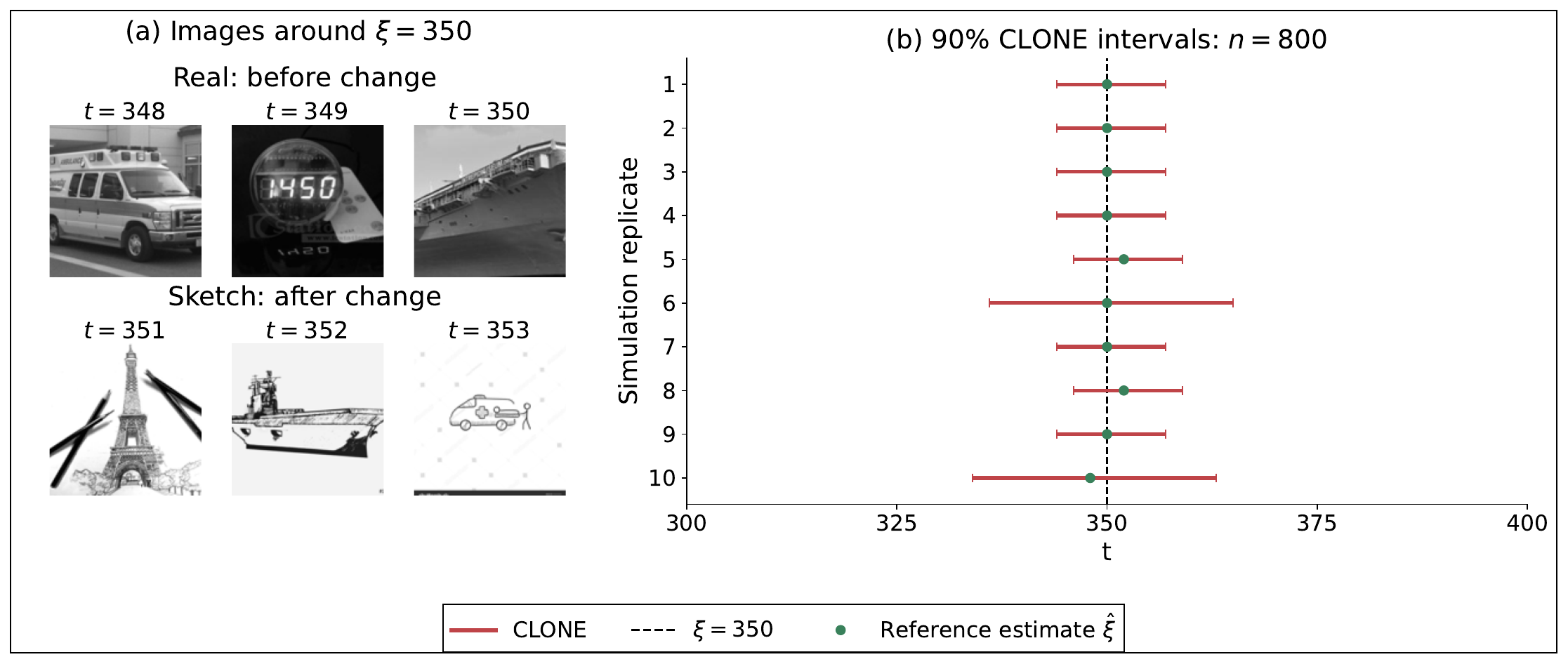}
    \caption{Domain shift localization: \clone{} gives contiguous confidence intervals around the reference estimates, shown in green.}
    \label{fig:domainnet-clone}
\end{figure}

Figure~\ref{fig:domainnet-clone} summarizes our findings. The left panel shows example images around the changepoint for one simulated sequence. The right panel plots the \clone{} intervals from the ten sequences, zoomed to the range $[300,400]$, with green points marking the corresponding reference estimates. The intervals remain narrow relative to the full timeline.

\subsection{SST-2: localizing sentiment change}\label{sec:sentiment_expt}
We next consider localizing sentiment changes using review snippets from the SST-2 dataset \citep{socher2013recursive}. We observe a sequence containing $n=1000$ reviews with a changepoint at $\xi=400$. Before the change, a review is positive with probability $(1+\delta)/2$; afterward, it is positive with probability $(1-\delta)/2$. We vary $\delta\in\{0.1,0.2,0.3,0.4,0.6,0.8,1\}$ and sample reviews independently with replacement from fixed positive and negative pools. Thus, $\delta$ measures the difference in positive-review proportions between the segments. Examples of positive and negative review snippets follow.
\begin{itemize}
    \item $t=399:$ ``juicy writer''
    \item $t=400:$ ``intricately structured and well-realized drama''
    \item $t=401:$ ``painfully''
    \item $t=402:$ ``than most of jaglom's self-conscious and gratingly irritating films''
\end{itemize}

We use a pretrained DistilBERT model fine-tuned for sentiment classification \citep{sanh2019distilbert}, with its logits providing the learned LLR-type score. We simulate $100$ sequences at each value of $\delta$, and compute the width of the \clone{} intervals.

\begin{figure}[t]
    \centering
    \includegraphics[width=0.85\linewidth]{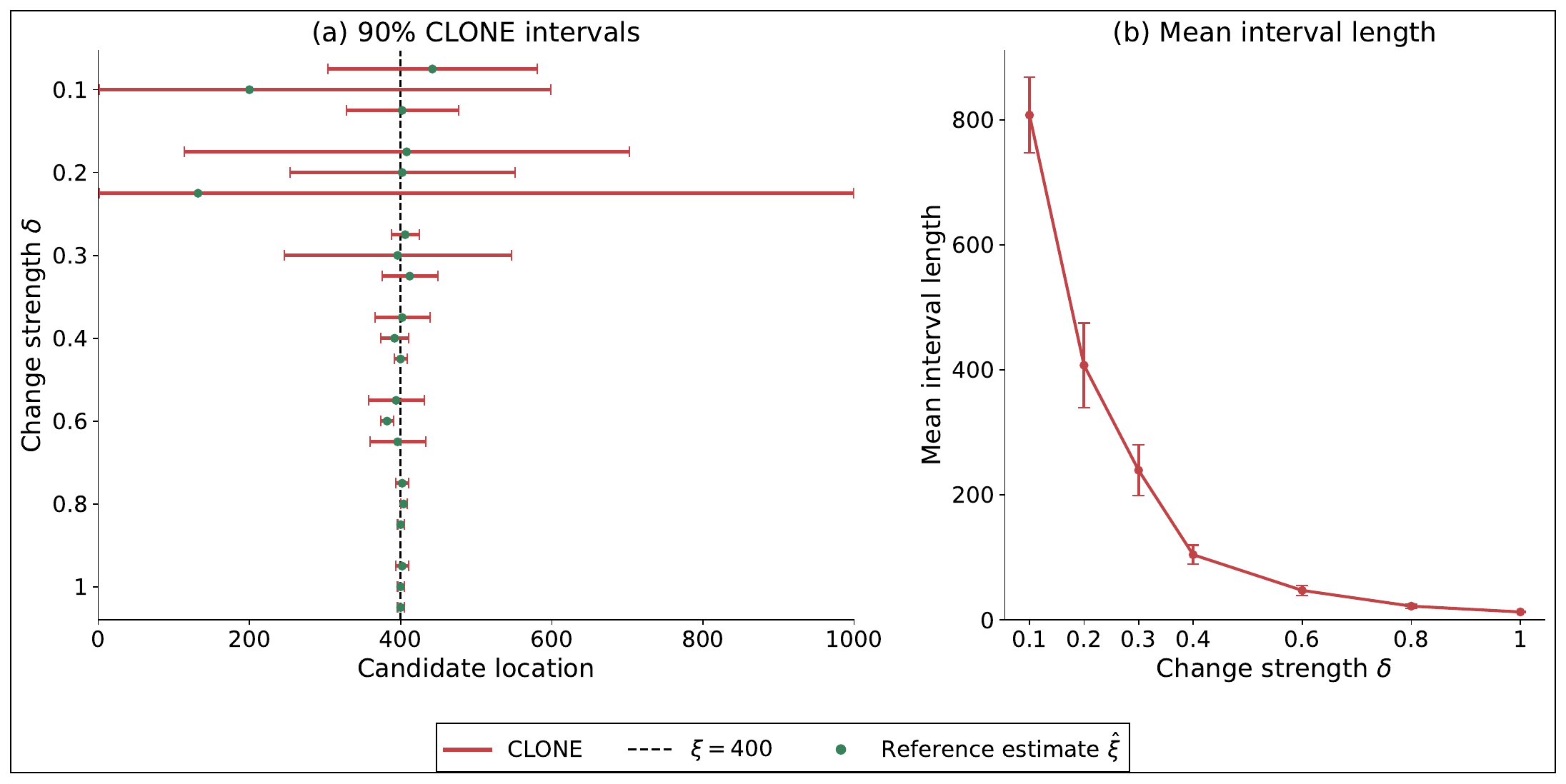}
    \caption{Sentiment shift localization: \clone{} intervals become narrower as sentiment separation increases, while quantifying uncertainty around the reference estimates shown in green.}
    \label{fig:sst2-clone}
\end{figure}

In the left panel of Figure~\ref{fig:sst2-clone}, we show \clone{} intervals for three simulated sequences at each $\delta$, with green points marking the corresponding reference estimates. As expected, if the initial reference estimate is far from $\xi$, the corresponding \clone{} interval tends to be wide. In the right panel, we plot the mean interval cardinality with $95\%$ Monte Carlo confidence intervals. The intervals are large when the sentiment shift is weak, i.e., $\delta$ is small.

\section{An alternative: conformal changepoint localization}\label{sec:conch}
The \conch{} framework of \citet{hore2026conformal} constructs distribution-free confidence sets by testing each candidate location separately and is a natural alternative for changepoint localization. A natural question is whether constructing intervals around the changepoint estimates reduces efficiency. We show that, with similarly constructed scores, \conch{} attains the same localization rate as \clone{}. When the oracle score $\ell$ is known, both methods attain the rate given by the information-theoretic lower bound on cardinality.

\subsection{The \conch{} construction}
The \conch{} framework requires the user to specify two components: a changepoint plausibility score and a split permutation group. A changepoint plausibility (CPP) score is any measurable map $S:\Xcal^n\times[n-1]\to\R$, where larger $S(\bX,t)$ indicates greater plausibility of a change at $t$.

Next, let $\mathcal S_n$ be the permutation group on $[n]$ and define $\Pi_t=\{\pi\in\mathcal S_n:\pi([t])=[t]\}$. These permutations act on the observations on either side of $t$ without mixing the segments. Writing $\pi(\bX)=(X_{\pi(1)},\ldots,X_{\pi(n)})$, \conch{} defines a conformal $p$-value for each $t\in[n-1]$ by
\begin{equation}\label{eq:conch}
 p_t=\frac{1}{|\Pi_t|}\sum_{\pi\in\Pi_t}
       \One{S(\pi(\bX),t)\le S(\bX,t)}.
\end{equation}
Under $\Hcal_{0,t}$, $p_t$ is a super-uniform random variable. Finally, one returns the time points where the \conch{} $p$-values exceed $\alpha$, that is, $\Ccal^{\conch}_{n,1-\alpha}=\{t\in[n-1]:p_t>\alpha\}$. Under Assumption~\ref{ass:exchangeability}, we therefore have
\begin{equation}\label{eq:conch-coverage}
 \P_\xi(\xi\in\Ccal^{\conch}_{n,1-\alpha})\ge1-\alpha.
\end{equation}
Further, we write $\bar\Ccal^{\conch}_{n,1-\alpha} =[\min\Ccal^{\conch}_{n,1-\alpha},\max\Ccal^{\conch}_{n,1-\alpha}] \cap[n-1]$ for the interval hull of the \conch{} confidence set, with $\bar\Ccal^{\conch}_{n,1-\alpha}=\varnothing$ when the set is empty. This hull also satisfies the same distribution-free coverage guarantee. 


In practice, sample splitting is recommended, as for the \clone{} confidence intervals. In particular, $D_{n,1}$ is used to learn the score, and $D_{n,2}$ is used for calibration. The original paper calls this implementation \splitconch{}. For simplicity, we refer to it as \conch{} here. For a formal description, see \cite[Section 3.3]{hore2026conformal}.


The authors further derive an expression for the optimal CPP score under the i.i.d.\ changepoint model and give practical recommendations for approximating it. These scores are similar to the LLR-type scores used for \clone{}. As in \clone{}, they use the training split $D_{n,1}$ to obtain an LLR estimate $\hat\ell_n:\Xcal\to\R$ and compute the reference estimate
\[
 \hat\xi_{n/2}\in\argmax_{s\in[m-1]}\sum_{i=1}^s\hat\ell_n(X_{2i-1,n}),
\]
and then define the learned LLR CPP score as
\begin{equation}\label{eq:score}
 S_n(\bY_n,t)=\sum_{i=1}^t\hat\ell_n(Y_{i,n})-\sum_{i=1}^{\hat\xi_{n/2}}\hat\ell_n(Y_{i,n}).
\end{equation}
This score is similar in spirit to the learned LLR window score, comparing the log-likelihood ratios of different segments.
As before, $\hat\ell_n$ may be obtained by estimating the two densities separately or by training a classifier and using its logits.

\subsection{Localization rates for the \conch{} interval hull}
\label{sec:split-sharpness}

We use the asymptotic setting and Assumption~\ref{ass:learned-score} from Section~\ref{sec:clone-sharpness}, including the independent dataset $D_n'$ and the reference estimate in~\eqref{eq:MLE}. The following theorem then characterizes the sharpness of both \conch{} and its interval hull under the generalized i.i.d.\ changepoint model.

\begin{theorem}
\label{thm:sharpness}
Suppose the model~\eqref{eq:iid-model} and Assumption~\ref{ass:learned-score} hold, with $r_n/n\to0$.
Then, for every fixed $\alpha\in(0,1)$,
\begin{equation}\label{eq:sharpness}
 \lim_{M\to\infty}\limsup_{n\to\infty}
 \P\Biggl(\,\,
 \max_{\substack{t\in[m-1]\\|t-\xi_n'|\ge Mr_n}}
 p_{t,n}>\alpha
 \Biggr)=0,
\end{equation}
where the maximum over an empty index set is interpreted as zero. Consequently, 
\begin{equation}\label{eq:learned-rate}
 |\bar\Ccal^{\conch}_{n,1-\alpha}|=\Op(r_n),
 \qquad
 \frac{|\bar\Ccal^{\conch}_{n,1-\alpha}|}{n-1}\xrightarrow{\P}0.
\end{equation}
\end{theorem}

Theorem~\ref{thm:sharpness} strengthens the asymptotic sharpness guarantee in \cite[Theorem~5.2]{hore2026conformal}. Their result shows that the normalized cardinality of the \conch{} confidence set vanishes, assuming that $\hat\ell_n$ consistently estimates the oracle LLR $\ell_n$. We instead give a precise localization rate for general LLR-type CPP scores. 

Therefore, \clone{} and the \conch{} interval hull attain the same localization rate. While \conch{} performs $O(n)$ separate permutation tests, \clone{} uses only $O(\log n)$ tests with geometric shrinking. 

\begin{figure}[t]
    \centering
    \includegraphics[width=\linewidth]{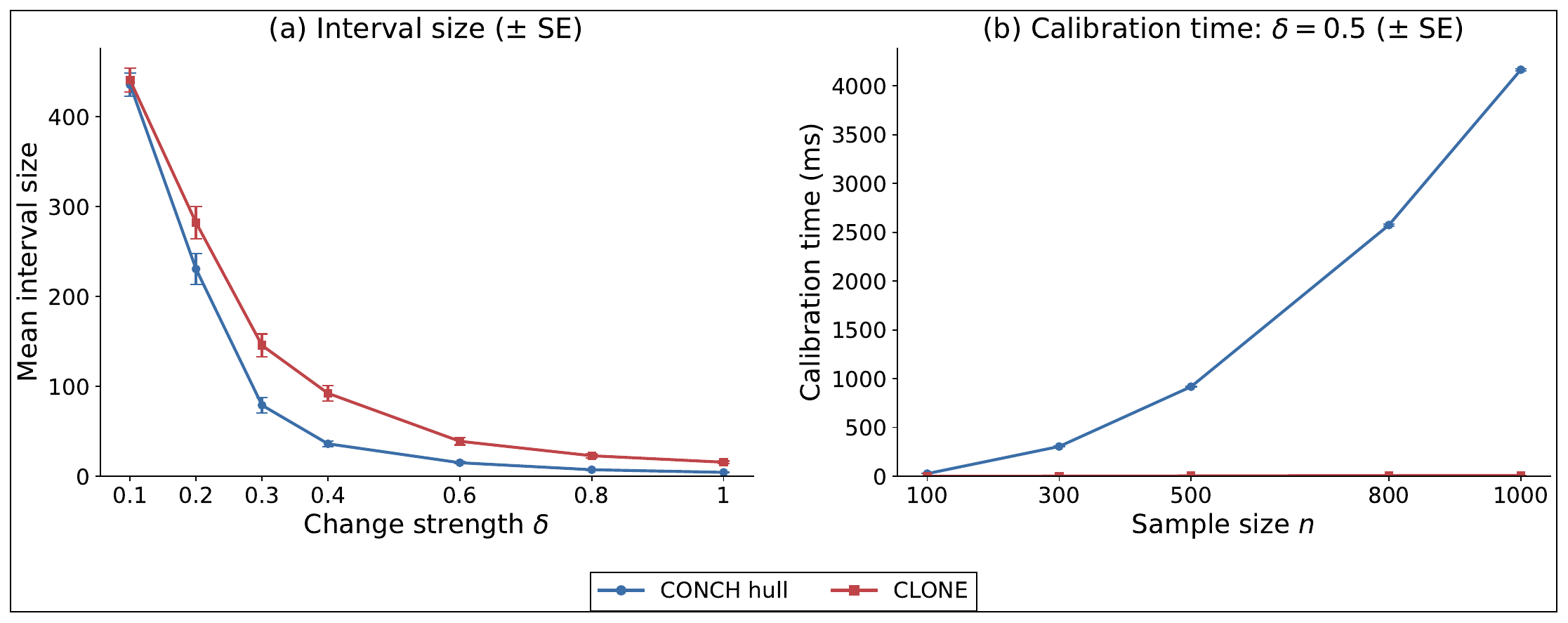}
    \caption{Gaussian mean shift: in finite samples, \clone{} intervals can be wider than the \conch{} hull but require substantially less calibration time.}
    \label{fig:conch-clone-comparison}
\end{figure}

\paragraph{Empirical comparison between \clone{} and \conch{}:}
We return to the Gaussian mean shift setting from Section~\ref{sec:clone-gaussian-simulation} and compare the interval width and calibration time of \clone{} and the \conch{} hull. As explained earlier, both methods use the same interlaced split and similar learned LLR scores. We therefore exclude the shared training stage and compare only the time spent on conformal calibration on $D_{n,2}$. We take $n=500$ and $\xi=250$ and vary the change strength $\delta$ as before to record the interval width. For calibration time, we fix $\delta=0.5$, set $\xi=n/2$, and vary $n$. Figure~\ref{fig:conch-clone-comparison} illustrates the tradeoff: the \conch{} hull gives narrower intervals in this experiment, whereas \clone{} requires substantially less calibration time. Matching localization rates do not imply equal finite-sample widths, and \clone{} offers a computationally efficient alternative.

\subsection{Oracle rates and information-theoretic optimality}
\label{sec:optimality}
We next specialize the sharpness results for \clone{} and the \conch{} hull to the oracle setting, where the oracle LLR $\hat\ell_n=\ell_n$ is known. This enables us to write down the localization rate $r_n$ in terms of the distributional gap between $P_{0,n}$ and $P_{1,n}$. 

We now define the symmetrized KL divergence, also known as Jeffreys divergence, by
\begin{equation}\label{eq:J}
 J_n =
 \KL(P_{0,n}\Vert P_{1,n})+
 \KL(P_{1,n}\Vert P_{0,n}).
\end{equation}
For the oracle LLR, $\mu_{0,n}=\KL(P_{0,n}\Vert P_{1,n})$ and $\mu_{1,n}=-\KL(P_{1,n}\Vert P_{0,n})$. Thus, the mean separation in condition~(ii) of Assumption~\ref{ass:learned-score} equals $J_n$, and the stability condition~(iii) of Assumption~\ref{ass:learned-score} holds immediately with $\varepsilon_n=0$.

\begin{corollary}
\label{cor:oracle}
Suppose $\xi_n/n\to\tau\in(0,1)$, both $\KL(P_{0,n}\Vert P_{1,n})$ and $ \KL(P_{1,n}\Vert P_{0,n})$ are positive and finite, and, for fixed $\kappa\in(0,1/2]$ and $C_v<\infty$,
\begin{equation}\label{eq:oracle-assumptions}
 \KL(P_{0,n}\Vert P_{1,n})\wedge
 \KL(P_{1,n}\Vert P_{0,n})
 \ge\kappa J_n,\qquad
 \max_{j=0,1}\Var_{P_{j,n}}(\ell_n(X))
 \le C_vJ_n.
\end{equation}
If $J_n\to0$ and $nJ_n\to\infty$, then, with $\hat\ell_n=\ell_n$, for every fixed $\alpha\in(0,1)$,
\begin{equation}\label{eq:oracle-fixed}
 | \hat\xi_{n/2}-\xi_n'|=\Op(J_n^{-1}),\quad
 |\bar\Ccal^{\splitconch}_{n,1-\alpha}|=\Op(J_n^{-1}),\quad
 |I^{\clone}_{n,1-\alpha}|=\Op(J_n^{-1}).
\end{equation}
\end{corollary}

Thus, in the oracle setting, both confidence-set constructions share the information-based localization scale $J_n^{-1}$. In particular, this rate cannot be improved by a valid distribution-free confidence set. 

For $t\in[n-1]$, let $\P_{t,n}=P_{0,n}^{\otimes t}\otimes P_{1,n}^{\otimes(n-t)}$. Fix $\alpha\in(0,1)$ and consider any distribution-free confidence set $\widetilde\Ccal_n\subseteq[n-1]$ satisfying
\begin{equation}\label{eq:honesty-fixed}
 \P_{t,n}(t\in\widetilde\Ccal_n)\ge1-\alpha
 \qquad\text{for every }t\in\mathcal{T}.
\end{equation}
We show that $\widetilde\Ccal_n$ cannot have cardinality $\mathrm{o}(J_n^{-1})$.

\begin{proposition}[A cardinality lower bound]\label{prop:lower-fixed}
Suppose $\xi_n/n\to\tau\in(0,1)$, the two  KL divergences $\KL(P_{0,n}\Vert P_{1,n})$ and $ \KL(P_{1,n}\Vert P_{0,n})$ are positive and finite, $J_n\to0$, and $nJ_n\to\infty$. For every fixed $\alpha\in(0,1)$, there exist constants $c_\alpha,p_\alpha>0$ such that every confidence set satisfying~\eqref{eq:honesty-fixed} obeys
\begin{equation}\label{eq:lower-fixed}
 \liminf_{n\to\infty}
 \P_{\xi_n,n}\left(|\widetilde\Ccal_n|\ge\frac{c_\alpha}{J_n}\right)
 \ge p_\alpha.
\end{equation}
In particular, $|\widetilde\Ccal_n|=\Omega_{P}(J_n^{-1})$.
\end{proposition}

Together, Corollary~\ref{cor:oracle} and Proposition~\ref{prop:lower-fixed} show that \clone{} and the \conch{} hull attain the optimal $J_n^{-1}$ localization rate at any fixed confidence level. 

\section{Discussion}\label{sec:discussion}
\clone{} provides a distribution-free confidence interval around a pre-learned changepoint estimate by testing exchangeability in nested intervals centered at that estimate. The distribution-free finite-sample validity of \clone{} allows flexibility in the choice of the preliminary estimate and window scores. With a geometrically shrinking radius grid, it requires only $O(\log n)$ permutation tests, providing a significant computational gain relative to the existing frameworks. The localization results show that this computational advantage preserves the asymptotic localization order. The fixed-level oracle bounds further show that \clone{} attains the rate given by the information-theoretic lower bound on cardinality. Extending this nested interval calibration approach to multiple changes and more general temporal dependence is a natural direction for future work.

\subsection*{Acknowledgment}
An AI model was used during the preparation of this manuscript to assist with language editing, presentation, and coding. The authors are responsible for reviewing the technical content and conclusions.

\bibliographystyle{chicago}
\bibliography{bibliography}

\clearpage
\appendix

\section{Post-detection changepoint localization}
\label{sec:selective_inference}

In the previous sections, we have focused on localizing a changepoint, assuming that a non-trivial change exists. We now study the possibility of no change. In this case, a natural goal is for the framework to return an empty set with high probability.

\clone{} framework offers a simple solution: first, we perform detection on the training split $D_{n,1}$. Let $A(D_{n,1})$ denote the resulting detection rule. We report the usual \clone{} interval when $A(D_{n,1})=1$ and the empty set otherwise. Since the validity of \clone{} established in  Theorem~\ref{thm:clone-valid} is proven conditional on $D_{n,1}$, even after this detection step, the coverage continues to hold. Consequently, whenever $\P_\xi(A(D_{n,1})=1)>0$,
\[
 \P_\xi\bigl(\xi\in I^{\clone}_{n,1-\alpha}\mid A(D_{n,1})=1\bigr)
 \ge 1-\alpha.
\]

If we instead want to use the whole sequence of observation $\bX$ for both detection and localization, modifying the full conformal implementation of \conch{} suitably can answer that. We write $\Hcal_{0,n}$ to denote the no-change hypothesis, $\xi=n$. \citet{hore2026conformal} note that the no-change hypothesis $\Hcal_{0,n}$ can be accommodated within the \conch{} framework by allowing $\xi=n$ and separately testing exchangeability of the whole sequence. In particular, writing $S(\cdot,n)$ for the plausibility score for no change, one can define the conformal $p$-value
\begin{equation}\label{eq:test_of_exch}
 p_n=\frac{1}{|\mathcal S_n|}
 \sum_{\pi\in\mathcal S_n}
 \One{S(\pi(\bX),n)\le S(\bX,n)}.
\end{equation}
The $p$-value $p_n$ is valid under $\Hcal_{0,n}$, as formally recorded below.
\begin{proposition}[Validity of the detection $p$-value]
\label{prop:detection-valid}
For any measurable score $S(\cdot,n)$, the $p$-value $p_n$ defined in~\eqref{eq:test_of_exch} satisfies $\P_{\Hcal_{0,n}}(p_n\le u)\le u$ for any $u\in(0,1)$. Consequently, for any detection level $\beta\in(0,1)$, the rule $A(\bX)=\One{p_n\le\beta}$ satisfies
\[
 \P_{\Hcal_{0,n}}(A(\bX)=1)
 \le\beta.
\]
\end{proposition}
One may therefore first test $\Hcal_{0,n}$ at a user-specified detection level $\beta\in(0,1)$ and return an empty set whenever $p_n>\beta$. Otherwise, if $p_n\le\beta$, one proceeds to localization. The \conch{} $p$-values $(p_t)_{t\in[n-1]}$, however, need not remain valid conditional on this detection event.

One may further ask whether a separate detection step is needed: would the original \conch{} set already be typically empty under $\Hcal_{0,n}$? The answer is no: the hypotheses $\Hcal_{0,t}$ specify independent, exchangeable segments but do not require their marginal distributions to differ. Under $\Hcal_{0,n}$, the law of $\bX$ is invariant under every permutation in $\Pi_t$ for each $t\in[n-1]$. Thus each $p_t$ in \eqref{eq:conch} remains super-uniform, even though there is no changepoint to localize. In particular, for any CPP score,
\begin{equation*}
 \E_{\Hcal_{0,n}}
 \bigl|\Ccal^{\conch}_{n,1-\alpha}\bigr|
 =
 \sum_{t=1}^{n-1}\P_{\Hcal_{0,n}}(p_t>\alpha)
 \ge(1-\alpha)(n-1).
\end{equation*}
The \conch{} confidence set therefore has expected size of order $n$ under no change. This is consistent with the $(n-1)$ hypotheses that \conch{} tests, but differs from the interpretation of an empty output as no detected change. A two-step procedure that reports coverage conditional on detection must therefore use conditionally valid calibration. We next describe how to modify \conch{} to achieve this.

\subsection{Detect-then-localize \conch{}}

Let $A:\Xcal^n\to\{0,1\}$ be any measurable changepoint detection rule, where $A(\bX)=1$ means that a change has been detected. For instance, the full-permutation $p$-value in \eqref{eq:test_of_exch} gives the rule $A(\bx)=\One{p_n(\bx)\le\beta}$ for a detection level $\beta\in(0,1)$. In general, any measurable detection rule may be used; a level-$\beta$ rule additionally controls false reports under no change.

Suppose a change is detected in an observed sequence $\bx$, so that $A(\bx)=1$. Define
\begin{equation}\label{eq:selective-pvalue}
 p_t^{\mathrm{det}}(\bx)=
 \frac{\displaystyle\sum_{\pi\in\Pi_t}
   \One{A(\pi(\bx))=1}
   \One{S(\pi(\bx),t)\le S(\bx,t)}}
 {\displaystyle\sum_{\pi\in\Pi_t}
   \One{A(\pi(\bx))=1}}.
\end{equation}
The resulting procedure, \detectconch{}, returns
\begin{equation*}
 \Ccal^{\detectconch}_{n,1-\alpha}=
 \begin{cases}
 \varnothing,&A(\bX)=0,\\
 \{t\in[n-1]:p_t^{\mathrm{det}}(\bX)>\alpha\},&A(\bX)=1.
 \end{cases}
\end{equation*}
Algorithm~\ref{alg:selective} records the full-permutation version of this procedure. On $A(\bx)=1$, the denominator is positive because the identity permutation is retained. Although the retained permutations need not form a group, the $p$-values remain valid conditional on detection under their corresponding null hypotheses.

\begin{theorem}\label{thm:selective}
Fix $\alpha\in(0,1)$. Suppose Assumption~\ref{ass:exchangeability} holds at a changepoint $\xi\in[n-1]$. Then $p_\xi^{\mathrm{det}}(\bX)$ is super-uniform conditional on $A(\bX)=1$, whenever the conditioning event has
positive probability. In particular,
\begin{equation}\label{eq:selective-coverage}
 \P_\xi\bigl(A(\bX)=1,\,
       \xi\notin \Ccal^{\detectconch}_{n,1-\alpha}\bigr)
 \le\alpha\,\P_\xi(A(\bX)=1).
\end{equation}
Consequently, whenever $\P_\xi(A(\bX)=1)>0$, we have $\P_\xi(\xi\in \Ccal^{\detectconch}_{n,1-\alpha} \mid A(\bX)=1)\ge1-\alpha$.
\end{theorem}

\begin{algorithm}[t]
\caption{\detectconch{}: localization conditional on detection}
\label{alg:selective}
\KwIn{Data $\bX$; detection rule $A$; CPP score $S$;
  confidence level $\alpha$.}
\KwOut{A changepoint confidence set.}
\If{$A(\bX)=0$}{\Return{$\varnothing$}\;}
\For{$t\in[n-1]$}{
  Retain $\pi\in\Pi_t$ for which $A(\pi(\bX))=1$\;
  Evaluate $S(\pi(\bX),t)$ on every retained permutation\;
  Compute $p_t^{\mathrm{det}}$ using
  \eqref{eq:selective-pvalue}\;
}
\Return{$\{t\in[n-1]:p_t^{\mathrm{det}}>\alpha\}$}\;
\end{algorithm}

\subsection{Numerical simulations}
\label{sec:det-conch-simulation}

We now evaluate \detectconch{} in a Gaussian mean shift setting. Specifically, we observe independent $X_1,\ldots,X_n$ with $X_i\sim\mathcal N(\delta,1)$ for $i\le\xi$ and $X_i\sim\mathcal N(-\delta,1)$ for $i>\xi$. Here, $\delta=0$ corresponds to no change, and for $\delta>0$, the changepoint is $\xi=n/2$. Throughout, we use detection level $\beta=0.05$ and localization level $\alpha=0.10$.

We use interlaced sample splitting: the odd subsequence is used to obtain a reference changepoint $\hat\xi_{n/2}$ by maximizing the absolute standardized CUSUM statistic, and to estimate Gaussian segment means and a common variance. The resulting learned LLR defines the CPP score on the even subsequence. For detection, we use the negative learned LLR partial sum at the reference estimate as the no-change plausibility score, and take $p_n\le\beta$ as the detection event.

\begin{figure}[t]
    \centering
    \includegraphics[width=\linewidth]{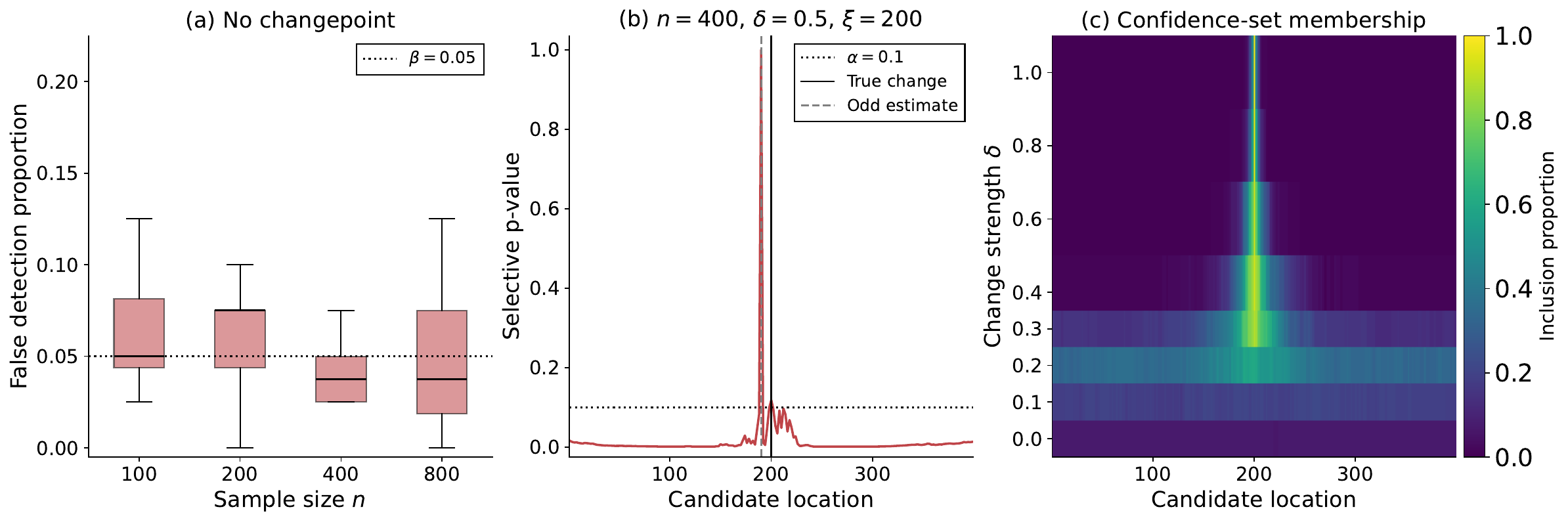}
    \caption{\detectconch{} controls false detections under no change. When a change is detected, stronger signals yield more concentrated confidence sets.}
    \label{fig:dtl-conch-gaussian}
\end{figure}

Figure~\ref{fig:dtl-conch-gaussian} summarizes our findings. In the left panel, we show boxplots of the empirical false detection proportion across sample sizes $n\in\{100,200,400,800\}$. The boxplots are centered near $\beta=0.05$, consistent with the theoretical bound on the false detection probability. In the middle panel, we plot the \detectconch{} $p$-values computed from the even subsequence on the original timeline, for $n=400$ and $\delta=0.5$. As expected, only locations near the true changepoint have $p$-values exceeding $\alpha$ and are retained.

Finally, the rightmost panel reports empirical confidence-set membership for $n=400$ and $\delta\in\{0,0.1,0.2,0.3,0.4,0.6,0.8,1\}$, using $100$ independent sequences at each signal strength. As $\delta$ increases, detection becomes more frequent and the confidence set concentrates around the true location. 

\section{Proofs}\label{app:proofs}
\subsection{Proving validity of \clone{}}\label{app:clone-validity}
\subsubsection{Proof of Theorem~\ref{thm:clone-valid}}
Throughout the proof, we condition on $D_{n,1}$, so that the reference estimate $\hat\xi_{n/2}$ and radius grid $R_0>\cdots>R_J$ are fixed. Let $\bY_{n,R}=(Y_{L_R,n},\ldots,Y_{U_R,n})$. If $\xi'\notin I_R$, then we must have $\xi'<L_R$ or $U_R\le\xi'$. In either case, by Assumption~\ref{ass:exchangeability}, the vector $\bY_{n,R}$ is exchangeable. Therefore,  conditional on $D_{n,1}$, $\pi(\bY_{n,R})\overset{d}{=}\bY_{n,R}$ for every $\pi\in\mathcal S_{k_R}$.

For every $\bx\in\Xcal^{k_R}$ and for $\pi\in\mathcal S_{k_R}$, we can write 
\[
p_{\mathrm{window},R}(\pi(\bx))=
 \frac1{|\mathcal S_{k_R}|}
 \sum_{\sigma\in\mathcal S_{k_R}}
 \One{S_R(\sigma(\bx))\ge S_R(\pi(\bx))}
\]
Hence, by \cite[Lemma~3]{harrison2012conservative}, we have
\begin{align*}
 \frac{1}{|\mathcal S_{k_R}|}\sum_{\pi\in\mathcal S_{k_R}}
 \One{p_{\mathrm{window},R}(\pi(\bx))\le u}
 &\le u,
 \qquad u\in(0,1).
\end{align*}
Consequently, it follows that
\begin{align*}
 \P_\xi(p_{\mathrm{window},R}\le u\mid D_{n,1})
 &=
 \frac1{|\mathcal S_{k_R}|}
 \sum_{\pi\in\mathcal S_{k_R}}
 \P_\xi\left(
 p_{\mathrm{window},R}(\pi(\bY_{n,R}))\le u
 \,\middle|\,D_{n,1}
 \right)\\
 &=
 \E_\xi\left[
 \frac1{|\mathcal S_{k_R}|}
 \sum_{\pi\in\mathcal S_{k_R}}
 \One{p_{\mathrm{window},R}(\pi(\bY_{n,R}))\le u}
 \,\middle|\,D_{n,1}
 \right]
 \le u.
\end{align*}

By definition,
$\hat j_\alpha=\min(\{j\in[J]:
p_{\mathrm{window},R_j}>\alpha\}\cup\{J+1\})$
and on $D_{n,2}$, we return $I_{R_{\hat j_\alpha-1}}$. If $\xi'\in I_{R_J}$, nestedness of the intervals immediately gives $\xi'\in I_{R_{\hat j_\alpha-1}}$.
Otherwise, define
\[j_*=\min\{j\in[J]:\xi'\notin I_{R_j}\}.\]
Since $I_{R_0}$ contains every calibration time point,
$j_*\ge1$, and
\begin{align*}
 \{\xi'\notin I_{R_{\hat j_\alpha-1}}\}=
 \{\hat j_\alpha>j_*\}=
 \bigcap_{j=1}^{j_*}
 \{p_{\mathrm{window},R_j}\le\alpha\}
 \subseteq
 \{p_{\mathrm{window},R_{j_*}}\le\alpha\}.
\end{align*}
The index $j_*$ is fixed conditional on $D_{n,1}$ and
$\xi'\notin I_{R_{j_*}}$. Hence,
\[
 \P_\xi(\xi'\notin I_{R_{\hat j_\alpha-1}}\mid D_{n,1})
 \le
 \P_\xi(p_{\mathrm{window},R_{j_*}}\le\alpha\mid D_{n,1})
 \le\alpha.
\]
Taking expectations and using $\xi\in B_{\xi'}$
proves \eqref{eq:clone-valid}.
\hfill $\square$

\subsubsection{Proof of Corollary~\ref{cor:clone-mc-valid}}
The proof follows the same approach as in the proof of Theorem~\ref{thm:clone-valid}. Fix any $R\in \N\cup\{0\}$. Under the null $\Hcal_{0,I_R}$, conditional on $D_{n,1}$ and the unordered window $\{Y_{L_R,n},\ldots,Y_{U_R,n}\}$, the observed score $S_R(Y_{L_R,n},\ldots,Y_{U_R,n})$ and its $B$ randomly permuted counterparts are exchangeable. Thus, following the arguments in the proof of Theorem~\ref{thm:clone-valid}, for any $u\in(0,1)$,
\[
 \P(\hat p_{\mathrm{window},R}\le u\mid D_{n,1})
 \le
 \frac{\lfloor(B+1)u\rfloor}{B+1}
 \le u.
\]
Here the probability is taken over both the randomness of the sampled permutations and the data sequence $\bX$.

Writing $j_*$ for the first index for which $\xi'\notin I_{R_{j_*}}$, we obtain
\[
 \P(\xi'\notin I_{R_{\hat j_\alpha-1}}\mid D_{n,1})
 \le
 \P(\hat p_{\mathrm{window},R_{j_*}}\le\alpha\mid D_{n,1})
 \le\alpha.
\]
Taking expectations and mapping back to the full timeline through the sets $B_t$ completes the proof.
\hfill$\square$

For the randomized versions, conditional on the unordered window, the randomized rank is uniform under exchangeability. The same holds for the Monte Carlo version by exchangeability of the $B+1$ scores, with ties randomized independently. Replacing the super-uniform rank bound above by this uniform rank bound gives the same first-excluded-interval argument.

\subsection{Proof of Theorem~\ref{thm:window-np}: Optimal score of \clone{}}

Throughout the proof, we condition on $D_{n,1}$ and the unordered observations in the window $\{Y_{L_R,n},\ldots,Y_{U_R,n}\}$. Thus, the reference estimate $\hat\xi_{n/2}$ and radius sequence are fixed. Fix $R$ with $\xi'\in I_R$.

Let $Q_0$ and $Q_\xi$ denote the laws of $(Y_{L_R,n},\ldots,Y_{U_R,n})$, conditional on $D_{n,1}$ and $\{Y_{L_R,n},\ldots,Y_{U_R,n}\}$, under the exchangeability $\Hcal_{0,I_R}$ and the alternative $\Hcal_{0,\xi}$ respectively. Since under $\Hcal_{0,I_R}$, every ordering is equally likely, $Q_0$ is essentially discrete uniform. Moreover, for an ordering $y=(y_{L_R},\ldots,y_{U_R})$,
\begin{align*}
 \frac{dQ_\xi}{dQ_0}(y)
 &\propto
 \prod_{i=L_R}^{\xi'}f_0(y_i)
 \prod_{i=\xi'+1}^{U_R}f_1(y_i)\\
 &=
 \left\{\prod_{i=L_R}^{U_R}f_1(y_i)\right\}
 \exp\left\{\sum_{i=L_R}^{\xi'}\ell(y_i)\right\}
 \propto \exp\{S_R^\star(y)\},
\end{align*}
where an empty product is interpreted as one. The last proportionality holds since the term $\left\{\prod_{i=L_R}^{U_R}f_1(y_i)\right\}$ is invariant to the ordering.

By the Neyman--Pearson lemma, therefore a most powerful conditional level-$\alpha$ test rejects when $S_R^\star>c$ and randomizes when $S_R^\star=c$, with rejection probability
$\gamma\in[0,1]$ chosen so that $Q_0(S_R^\star>c)+\gamma Q_0(S_R^\star=c)=\alpha$.
Equivalently, sampling $V_R\sim\mathrm{Unif}(0,1)$ independent of the data, for any score $S_R$, its randomized
permutation $p$-value can be given by
\begin{align*}
   p_{\mathrm{window},R}(S_R;y)
 &= Q_0(S_R>S_R(y)) +V_R Q_0(S_R=S_R(y))\\
 &= \frac1{|\mathcal S_{k_R}|}
 \sum_{\pi\in\mathcal S_{k_R}}
 \One{
 S_R(\pi(y_{L_R},\ldots,y_{U_R}))
 > S_R(y_{L_R},\ldots,y_{U_R})}\\
 &\hspace{1.7cm} + \frac{V_R}{|\mathcal S_{k_R}|}
 \sum_{\pi\in\mathcal S_{k_R}}
 \One{
 S_R(\pi(y_{L_R},\ldots,y_{U_R}))
 = S_R(y_{L_R},\ldots,y_{U_R})}.
\end{align*}
Consequently, from the above discussion, we obtain
\[
 Q_\xi\bigl(p_{\mathrm{window},R}(S_R)\le\alpha\bigr)
 \le
 Q_\xi\bigl(p_{\mathrm{window},R}(S_R^\star)\le\alpha\bigr).
\]
This proves the first part.

For the second part, since $j_\star$ is fixed conditional on $D_{n,1}$ and $\xi'\in I_{R_j}$ for every $j\le j_\star$, averaging the preceding power inequality over $\{Y_{L_R,n},\ldots,Y_{U_R,n}\}$ gives
\[
 \P_\xi\left(
 p_{\mathrm{window},R_j}(S_{R_j})\le\alpha
 \,\middle|\,D_{n,1}
 \right)
 \le
 \P_\xi\left(
 p_{\mathrm{window},R_j}(S^\star_{R_j})\le\alpha
 \,\middle|\,D_{n,1}
 \right).
\]
Consequently, by linearity of conditional expectation,
\begin{align*}
 \E_\xi\left[
 \sum_{j=1}^{j_\star}
 \One{p_{\mathrm{window},R_j}(S_{R_j})\le\alpha}
 \,\middle|\,D_{n,1}
 \right]
 &=
 \sum_{j=1}^{j_\star}
 \P_\xi\left(
 p_{\mathrm{window},R_j}(S_{R_j})\le\alpha
 \,\middle|\,D_{n,1}
 \right)\\
 &\le
 \sum_{j=1}^{j_\star}
 \P_\xi\left(
 p_{\mathrm{window},R_j}(S^\star_{R_j})\le\alpha
 \,\middle|\,D_{n,1}
 \right)\\
 &=
 \E_\xi\left[
 \sum_{j=1}^{j_\star}
 \One{p_{\mathrm{window},R_j}(S^\star_{R_j})\le\alpha}
 \,\middle|\,D_{n,1}
 \right],
\end{align*}
which proves the second part.

\hfill $\square$

\subsection{Localization of the reference estimate}\label{app:reference-rate}
\subsubsection{Proof of Proposition~\ref{prop:MLE}}

We start by defining the event
\[
 G_n=
 \left\{
 \E_{P_{0,n}}[\hat\ell_n(X)\mid D_n']
 \ge\mu_{0,n}/2,\quad
 \E_{P_{1,n}}[\hat\ell_n(X)\mid D_n']
 \le\mu_{1,n}/2
 \right\}.
\]
By Chebyshev's inequality, we obtain
\[
 \P\left(
 \E_{P_{0,n}}[\hat\ell_n(X)\mid D_n']<\mu_{0,n}/2
 \right)
 \le
 \frac{
 4\Var_{D_n'}\!\left(
 \E_{P_{0,n}}[\hat\ell_n(X)\mid D_n']
 \right)}
 {\mu_{0,n}^2},
\]
and similarly,
\[
 \P\left(
 \E_{P_{1,n}}[\hat\ell_n(X)\mid D_n']>\mu_{1,n}/2
 \right)
 \le
 \frac{
 4\Var_{D_n'}\!\left(
 \E_{P_{1,n}}[\hat\ell_n(X)\mid D_n']
 \right)}
 {\mu_{1,n}^2}.
\]
Consequently, by Assumption~\ref{ass:learned-score}(ii)
and the definition of $\varepsilon_n$,
\begin{align}\label{eq:G-prob}
 \P(G_n^c)
 &\le
 4\left\{
 \frac{
 \Var_{D_n'}\!\left(
 \E_{P_{0,n}}[\hat\ell_n(X)\mid D_n']
 \right)}
 {\mu_{0,n}^2}
 +
 \frac{
 \Var_{D_n'}\!\left(
 \E_{P_{1,n}}[\hat\ell_n(X)\mid D_n']
 \right)}
 {\mu_{1,n}^2}
 \right\}
 \le\frac{4\varepsilon_n}{\kappa^2}.
\end{align}
For sufficiently large $n$, Assumption~\ref{ass:learned-score}(i)
gives $\xi_n''\in[m-1]$. Thus, for every integer $L\ge1$,
\[
 \{\hat\xi_{n/2}\ge\xi_n''+L\}
 \subseteq
 \left\{
 \max_{L\le l\le m-1-\xi_n''}
 \sum_{i=\xi_n''+1}^{\xi_n''+l}
 \hat\ell_n(X_{2i-1,n})\ge0
 \right\}.
\]
Conditional on $D_n'$, the observations in this sum are
independent draws from $P_{1,n}$, and on $G_n$, each
summand has conditional mean at most $\mu_{1,n}/2$.
Therefore, on $G_n$, the first bound in
Lemma~\ref{lem:maximal} gives
\begin{align*}
 &\P(\hat\xi_{n/2}\ge\xi_n''+L\mid D_n')\\
 &\quad\le
 \P\left(
 \max_{L\le l\le m-1-\xi_n''}
 \frac1l
 \sum_{i=\xi_n''+1}^{\xi_n''+l}
 \left\{
 \hat\ell_n(X_{2i-1,n})
 -
 \E_{P_{1,n}}[\hat\ell_n(X)\mid D_n']
 \right\}
 \ge-\mu_{1,n}/2
 \,\middle|\,D_n'
 \right)\\
 &\quad\le
 \frac{
 4\Var_{P_{1,n}}(\hat\ell_n(X)\mid D_n')}
 {L\mu_{1,n}^2}.
\end{align*}
On $G_n$, an analogous argument gives
\[
 \P(\hat\xi_{n/2}\le\xi_n''-L\mid D_n')
 \le
 \frac{
 4\Var_{P_{0,n}}(\hat\ell_n(X)\mid D_n')}
 {L\mu_{0,n}^2}.
\]
By the law of total variance,
\[
 \E\!\left[
 \Var_{P_{j,n}}(\hat\ell_n(X)\mid D_n')
 \right]
 \le\sigma_{j,n}^2,
 \qquad j\in\{0,1\}.
\]
Consequently, taking $L=u-1$ gives
\[
 \P(|\hat\xi_{n/2}-\xi_n'|>u)
 \le
 \P(G_n^c)
 +
 \frac4{u-1}
 \left(
 \frac{\sigma_{0,n}^2}{\mu_{0,n}^2}
 +
 \frac{\sigma_{1,n}^2}{\mu_{1,n}^2}
 \right),
\]
because $|\xi_n''-\xi_n'|\le1$. This intermediate bound
requires only $\mu_{0,n}>0>\mu_{1,n}$. By the balance
condition in Assumption~\ref{ass:learned-score}(ii),
\[
 \frac{\sigma_{0,n}^2}{\mu_{0,n}^2}
 +
 \frac{\sigma_{1,n}^2}{\mu_{1,n}^2}
 \le
 \frac{\sigma_{0,n}^2+\sigma_{1,n}^2}
 {\kappa^2(\mu_{0,n}-\mu_{1,n})^2}.
\]
Combining this with~\eqref{eq:G-prob} proves
\eqref{eq:MLE-tail}. For $u=\lfloor Mr_n\rfloor$ and $M\ge4$, the resulting
bound is at most
$4\varepsilon_n/\kappa^2+8/(\kappa^2M)$. Thus
\[
 \lim_{M\to\infty}\limsup_{n\to\infty}
 \P(|\hat\xi_{n/2}-\xi_n'|>Mr_n)=0.
\]
Finally,
$|2\hat\xi_{n/2}-\xi_n|
\le2|\hat\xi_{n/2}-\xi_n'|+1$, and $r_n\ge1$.
This proves the second part.
\hfill$\square$

\subsubsection{Auxiliary lemma}
\begin{lemma}[A maximal inequality]\label{lem:maximal}
Let $W_1,\ldots,W_N$ be independent, with $\E[W_i]=0$
and $\Var(W_i)\le V$ for every $i\in[N]$. For integers
$N\ge L\ge1$ and $x>0$,
\begin{equation}\label{eq:maximal}
 \P\left(
 \max_{L\le l\le N}
 \frac{|\sum_{i=1}^{l}W_i|}{l}\ge x
 \right)
 \le
 \begin{cases}
 \displaystyle\frac{V}{Lx^2},
 &\text{if the variables are i.i.d.},\\[1ex]
 \displaystyle\frac{4V}{Lx^2},
 &\text{in general}.
 \end{cases}
\end{equation}
\end{lemma}
\begin{proof}
    For the i.i.d.\ bound, see
\cite[Theorem~2.2]{ramdas2026randomized}.
The proof of the general result is as follows. 

We start by observing that by union bound,
\[
\P\left(\max_{L\le l\le N}
       \frac{|\sum_{i=1}^lW_i|}{l}\ge x\right)\le \sum_{\substack{j\ge0\\2^jL\le N}} \P\left(\max_{2^jL\le l\le \min\{2^{j+1}L,N\}}
       \frac{|\sum_{i=1}^lW_i|}{l}\ge x\right).
\]
Now, fix any $j\ge 0$, and note that for $2^jL\le l<2^{j+1}L$,
\[
\left\{\frac{|\sum_{i=1}^lW_i|}{l}\ge x\right\}\implies \left\{\bigl|\sum_{i=1}^lW_i\bigr|\ge2^jLx\right\}
\]
Consequently, Kolmogorov's inequality gives
\[
\P\left(\max_{1\le l\le\min\{2^{j+1}L,N\}}\bigl|\sum_{i=1}^lW_i\bigr|\ge2^jLx\right) \le\frac{2^{j+1}LV}{(2^jLx)^2}
 =\frac{2^{1-j}V}{Lx^2}.
\]
Consequently, we obtain:
\[
 \P\left(\max_{L\le l\le N}
       \frac{|\sum_{i=1}^lW_i|}{l}\ge x\right)\le \sum_{j=0}^\infty \frac{2^{1-j}V}{Lx^2}=\frac{4V}{Lx^2}.
\]
This completes the proof.
\end{proof}

\subsection{Proving sharpness of \clone{}}
\label{app:clone-sharpness}

\subsubsection{Proof of Theorem~\ref{thm:clone-sharp}}

We start by fixing $c>0$ such that $2c<\min\{\tau,1-\tau\}$. For all sufficiently large $n$, therefore, by Assumption~\ref{ass:learned-score}(i), we have $\xi_n'\in[2cm,(1-2c)m]$. Fix $M$ sufficiently large, and write $A=\lceil Mr_n\rceil$ and $u=\lfloor A/4\rfloor$. Since by the hypothesis we have $r_n=o(n)=o(m)$, eventually it holds that $A\le cm$.

Define
\[
 H_n=\left\{
 \E_{P_{0,n}}[\hat\ell_n(X)\mid D_n']
 -\E_{P_{1,n}}[\hat\ell_n(X)\mid D_n']
 \ge\frac{\mu_{0,n}-\mu_{1,n}}2
 \right\}.
\]
The conditional mean separation has mean $\mu_{0,n}-\mu_{1,n}$ and variance at most $2\varepsilon_n(\mu_{0,n}-\mu_{1,n})^2$. Chebyshev's inequality therefore gives $\P(H_n^c)\le8\varepsilon_n$. For the next steps, we condition on $(D_n',D_{n,1})$ and restrict to $H_n\cap\{|\hat\xi_{n/2}-\xi_n'|\le u\}$. Since $u\le A\le cm$, on this event $\hat\xi_{n/2}\in[cm,(1-c)m]$, and the radius sequence is fixed.

\paragraph{Step 1: True window scores are large.}
Fix a radius $R\ge A$. The lengths of the segments to the left and right of $\hat\xi_{n/2}$ are respectively $a_R=\min\{R+1,\hat\xi_{n/2}\}$ and  $b_R=\min\{R+1,m-\hat\xi_{n/2}\}$. By the above discussion, both are at least $A$.

If $\hat\xi_{n/2}\le\xi_n'$, then all the observations to the left of $\hat\xi_{n/2}$ are from the pre-change distribution, while exactly $\xi_n'-\hat\xi_{n/2}$ observations on the right are pre-change. Hence, the window score $S_R$ has the expectation,
\begin{align*}
 \E[S_R\mid D_n',D_{n,1}]=\left(1-\frac{\xi_n'-\hat\xi_{n/2}}{b_R}\right)\left\{\E_{P_{0,n}}[\hat\ell_n(X)\mid D_n']-\E_{P_{1,n}}[\hat\ell_n(X)\mid D_n']\right\}.
\end{align*}
If instead $\hat\xi_{n/2}\ge\xi_n'$, the same expression holds with the pre-factor replaced by $1-(\hat\xi_{n/2}-\xi_n')/a_R$. Since $u/A\le1/4$, by the above calculation, on $H_n$ we have
\begin{equation}\label{eq:window-drift}
 \E[S_R\mid D_n',D_{n,1}]
 \ge\frac38(\mu_{0,n}-\mu_{1,n}).
\end{equation}

Next, let us write $V_n=\max_{j=0,1} \Var_{P_{j,n}}(\hat\ell_n(X)\mid D_n')$. Since the calibration observations are drawn independently, we further have $\Var(S_R\mid D_n',D_{n,1}) \le V_n(a_R^{-1}+b_R^{-1})$. Therefore, by Chebyshev's inequality, we obtain
\begin{equation}\label{eq:window-observed-tail}
 \P\left(
 S_R<\frac{\mu_{0,n}-\mu_{1,n}}4
 \,\middle|\,D_n',D_{n,1}
 \right)
 \le
 \frac{64V_n}{(\mu_{0,n}-\mu_{1,n})^2}
 \left(\frac1{a_R}+\frac1{b_R}\right).
\end{equation}

\paragraph{Step 2: Control the permutation tail.}
Let $k_R=a_R+b_R$ be the length of the window segment and $\bar\ell_R=k_R^{-1} \sum_{i=L_R}^{U_R}\hat\ell_n(Y_{i,n})$ be the empirical average of $\hat{\ell}$ computed on the corresponding observations. Further, note that conditional on the observed window, the permutation mean of $S_R$ is zero. This is because both permuted averages have conditional expectation $\bar\ell_R$. Moreover, we can compute
\begin{align}
 &\Var_\pi\left(
 S_R(\pi(Y_{L_R,n},\ldots,Y_{U_R,n}))
 \,\middle|\,D_n',D_{n,1},\bY_n
 \right)
 \nonumber\\
 &\qquad=
 \frac{k_R}{k_R-1}
 \left(\frac1{a_R}+\frac1{b_R}\right)
 \frac1{k_R}
 \sum_{i=L_R}^{U_R}
 \bigl(\hat\ell_n(Y_{i,n})-\bar\ell_R\bigr)^2
 \nonumber\\
 &\qquad\le
 2\left(\frac1{a_R}+\frac1{b_R}\right)
 \frac1{k_R}
 \sum_{i=L_R}^{U_R}
 \left(
 \hat\ell_n(Y_{i,n})
 -
 \frac{\mu_{0,n}+\mu_{1,n}}2
 \right)^2,
 \label{eq:window-permutation-var}
\end{align}
where the last inequality uses the fact that the empirical mean $\bar\ell_R$ minimizes squared deviations and $k_R\ge2$. On the event $\{S_R\ge(\mu_{0,n}-\mu_{1,n})/4\}$, Chebyshev's inequality under the permutation law gives the following upper bound on the \clone{} $p$-values:
\[ 
 p_{\mathrm{window},R}
 \le
 \frac{16}{(\mu_{0,n}-\mu_{1,n})^2}
 \Var_\pi\left(
 S_R(\pi(Y_{L_R,n},\ldots,Y_{U_R,n}))
 \,\middle|\,D_n',D_{n,1},\bY_n
 \right).
\]
For notational clarity, let us write
\[
 K_n=
 \E_{P_{0,n}}\left[
 \left(
 \hat\ell_n(X)-\frac{\mu_{0,n}+\mu_{1,n}}2
 \right)^2
 \,\middle|\,D_n'
 \right]+\E_{P_{1,n}}\left[
 \left(
 \hat\ell_n(X)-\frac{\mu_{0,n}+\mu_{1,n}}2
 \right)^2
 \,\middle|\,D_n'
 \right].
\]
The empirical second
moment in~\eqref{eq:window-permutation-var}, $\frac1{k_R}
 \sum_{i=L_R}^{U_R}
 \left(\hat\ell_n(Y_{i,n})-\frac{\mu_{0,n}+\mu_{1,n}}2
 \right)^2$ has expectation at most $K_n$, conditional on $D'_n, D_{n,1}$. Combining Markov's inequality with
\eqref{eq:window-observed-tail} therefore gives
\begin{align}\label{eq:window-nonrejection}
 &\P(p_{\mathrm{window},R}>\alpha
       \mid D_n',D_{n,1})\nonumber\\
 &\quad\le
 \P\left(
 S_R<\frac{\mu_{0,n}-\mu_{1,n}}4
 \,\middle|\,D_n',D_{n,1}
 \right)\nonumber\\
 &\qquad+
 \P\left(
 \frac1{k_R}\sum_{i=L_R}^{U_R}
 \left(\hat\ell_n(Y_{i,n})
       -\frac{\mu_{0,n}+\mu_{1,n}}2\right)^2
 >
 \frac{\alpha(\mu_{0,n}-\mu_{1,n})^2}
      {32(a_R^{-1}+b_R^{-1})}
 \,\middle|\,D_n',D_{n,1}
 \right)\nonumber\\
 &\quad\le
 \frac{64V_n+32K_n/\alpha}
      {(\mu_{0,n}-\mu_{1,n})^2}
 \left(\frac1{a_R}+\frac1{b_R}\right).
\end{align}
This bound holds on the event $H_n\cap\{|\hat\xi_{n/2}-\xi_n'|\le u\}$ for every radius $R\ge A$.

\paragraph{Step 3: Sum over the geometric radii.}
We first observe that, by the law of total variance and
the definitions of the unconditional moments,
\begin{align*}
 \E[V_n]
 \le \sigma_{0,n}^2+\sigma_{1,n}^2,
 \quad\E[K_n]=\sigma_{0,n}^2+\sigma_{1,n}^2
   +\frac{(\mu_{0,n}-\mu_{1,n})^2}{2}.
\end{align*}
Conditional on $(D_n',D_{n,1})$, the radius sequence is
fixed. On $H_n\cap\{|\hat\xi_{n/2}-\xi_n'|\le u\}$, a union
bound, followed by~\eqref{eq:window-nonrejection} and
Lemma~\ref{lem:geometric-sum}, therefore gives
\begin{align*}
 &\P\left(
 \exists j\in[J]:R_j\ge A,\,
 p_{\mathrm{window},R_j}>\alpha
 \,\middle|\,D_n',D_{n,1}
 \right)\\
 &\quad\le
 \frac{64V_n+32K_n/\alpha}
      {(\mu_{0,n}-\mu_{1,n})^2}
 \sum_{\substack{j\in[J]\\R_j\ge A}}
\left(\frac1{a_{R_j}}+\frac1{b_{R_j}}\right)\le
 \frac{4(64V_n+32K_n/\alpha)}
      {cA(\mu_{0,n}-\mu_{1,n})^2}.
\end{align*}
The aforementioned bound holds uniformly over $\hat\xi_{n/2}\in[cm,(1-c)m]$. Taking expectations
therefore yields
\begin{align*}
 &\P\left(
 \exists j\in[J]:R_j\ge A,\,
 p_{\mathrm{window},R_j}>\alpha,\,
 H_n,\ |\hat\xi_{n/2}-\xi_n'|\le u
 \right)\\
 &\quad\le
 \frac{4\{64\E[V_n]+32\E[K_n]/\alpha\}}
      {cA(\mu_{0,n}-\mu_{1,n})^2}\\
 &\quad\le
 \frac4{\alpha cA}
 \left\{
 (64\alpha+32)
 \frac{\sigma_{0,n}^2+\sigma_{1,n}^2}
      {(\mu_{0,n}-\mu_{1,n})^2}
 +16
 \right\}\le
 \frac{448r_n}{\alpha cA}
 \le\frac{448}{\alpha cM}.
\end{align*}
Here, the penultimate inequality uses $\alpha<1$,
$r_n\ge1$, and
$(\sigma_{0,n}^2+\sigma_{1,n}^2)/
(\mu_{0,n}-\mu_{1,n})^2\le r_n$.

\paragraph{Step 4: Bound the returned interval.}
Suppose every tested radius $R_j\ge A$ rejects.
If the first nonrejection occurs at index $j$, then
$R_j<A$, and the procedure returns $I_{R_{j-1}}$.
Since $R_j=\lfloor R_{j-1}/2\rfloor$, we have
$R_{j-1}\le2R_j+1\le2A+1$.
If all tests reject, the returned radius is zero.
Thus, in either case, the retained calibration interval
has at most $4A+3$ locations. Since each calibration
location maps to at most two original locations,
\[
 |I^{\clone}_{n,1-\alpha}|\le8A+6.
\]
Consequently, Step 3 and $\P(H_n^c)\le8\varepsilon_n$
give
\begin{align*}
 \P\left(
 |I^{\clone}_{n,1-\alpha}|>8A+6
 \right)
 &\le
 8\varepsilon_n
 +\P(|\hat\xi_{n/2}-\xi_n'|>u)
 +\frac{448}{\alpha cM}.
\end{align*}

For $M\ge16$, we have $A\ge16$ and
$u-1=\lfloor A/4\rfloor-1\ge A/8$.
Proposition~\ref{prop:MLE} therefore implies
\begin{align*}
 \P(|\hat\xi_{n/2}-\xi_n'|>u)
 &\le
 \frac{4\varepsilon_n}{\kappa^2}
 +\frac{4}{\kappa^2(u-1)}
 \frac{\sigma_{0,n}^2+\sigma_{1,n}^2}
      {(\mu_{0,n}-\mu_{1,n})^2}\\
 &\le
 \frac{4\varepsilon_n}{\kappa^2}
 +\frac{32r_n}{\kappa^2A}
 \le
 \frac{4\varepsilon_n}{\kappa^2}
 +\frac{32}{\kappa^2M}.
\end{align*}
Combining these bounds, for every fixed $M\ge16$ and
all sufficiently large $n$,
\[
 \P\left(
 |I^{\clone}_{n,1-\alpha}|>8\lceil Mr_n\rceil+6
 \right)
 \le
 \left(8+\frac4{\kappa^2}\right)\varepsilon_n
 +\frac{32}{\kappa^2M}
 +\frac{448}{\alpha cM}.
\]
By Assumption~\ref{ass:learned-score}(iii),
$\varepsilon_n\to0$. First letting $n\to\infty$ and
then $M\to\infty$ proves
$|I^{\clone}_{n,1-\alpha}|=\Op(r_n)$.
The relative size converges to zero because $r_n/n\to0$.
\hfill$\square$

\subsubsection{Auxiliary lemma}

\begin{lemma}
\label{lem:geometric-sum}
Fix $c\in(0,1/2)$ and an integer reference
$h\in[m-1]\cap[cm,(1-c)m]$. Consider the geometric grid defined by $R_0=\max\{h,m-1-h\}$ and $R_{j+1}=\lfloor R_j/2\rfloor$ until $R_J=0$.
Write $a_R=\min\{R+1,h\}$ and
$b_R=\min\{R+1,m-h\}$.
Then, for every integer $1\le A\le cm$,
\begin{equation}\label{eq:geometric-sum}
 \sum_{\substack{j\in[J]\\R_j\ge A}}
 \left(\frac1{a_{R_j}}+\frac1{b_{R_j}}\right)
 \le\frac4{cA}.
\end{equation}
\end{lemma}

\begin{proof}
Since $R_0\le m-1$, every radius satisfies $R+1\le m$.
Together with $h\wedge(m-h)\ge cm$, this gives
\[
 a_R\wedge b_R
 \ge\min\{R+1,cm\}
 \ge c(R+1).
\]
If no tested radius is at least $A$, the claim is
immediate. Otherwise, let
$j_A=\max\{j\in[J]:R_j\ge A\}$.
The recursion gives $R_j\ge2R_{j+1}$, so
$R_{j_A-k}\ge2^kR_{j_A}\ge2^kA$
for $k=0,\ldots,j_A-1$. Therefore,
\begin{align*}
 \sum_{\substack{j\in[J]\\R_j\ge A}}
 \left(\frac1{a_{R_j}}+\frac1{b_{R_j}}\right)
\le
 \frac2c\sum_{j=1}^{j_A}\frac1{R_j+1}\le
 \frac2{cA}\sum_{k=0}^{j_A-1}2^{-k}
 \le\frac4{cA}.
\end{align*}
This proves the claim.
\end{proof}
\subsection{Proving sharpness of \conch}
We start by defining some notation for clarity of the exposition. For $t\in[m-1]$, define
\[
 \hat\mu_{t,n,L}
 =\frac1t\sum_{i=1}^{t}\hat\ell_n(Y_{i,n}),
 \qquad
 \hat\mu_{t,n,R}
 =-\frac1{m-t}\sum_{i=t+1}^{m}\hat\ell_n(Y_{i,n}).
\]
For $t\ne\hat\xi_{n/2}$, write
\begin{equation}\label{eq:split-score-gap}
 \Delta_{t,\hat\xi_{n/2}}
 =
 \frac{S_n(\bY_n,t)}{|t-\hat\xi_{n/2}|}
 -
 \hat\mu_{t,n,L}\One{t>\hat\xi_{n/2}}
 -
 \hat\mu_{t,n,R}\One{t<\hat\xi_{n/2}}.
\end{equation}
We also define
\begin{equation}\label{eq:split-second-moment}
 \hat v_n
 =
 \frac1m\sum_{i=1}^{m}
 \left(
 \hat\ell_n(Y_{i,n})
 -
 \frac{\mu_{0,n}+\mu_{1,n}}2
 \right)^2.
\end{equation}
\label{app:split-sharpness}

\subsubsection{Proof of Theorem~\ref{thm:sharpness}}

Fix $c\in(0,\min\{\tau,1-\tau\})$. For all sufficiently
large $n$, Assumption~\ref{ass:learned-score}(i) gives
\[
 \xi_n'\ge cm,\qquad m-\xi_n'\ge cm.
\]
Throughout the proof, write $R=\lceil Mr_n\rceil$
and $u=\lfloor cR/(2+c)\rfloor$. Then $R>u$ and
\begin{equation}\label{eq:R-u}
 \frac{u}{R-u}\le\frac c2.
\end{equation}

\paragraph{Step 1: Restrict to a high-probability event.}
We start by defining the event
\[
 H_n=
 \left\{
 \E_{P_{0,n}}[\hat\ell_n(X)\mid D_n']
 -
 \E_{P_{1,n}}[\hat\ell_n(X)\mid D_n']
 \ge\frac{\mu_{0,n}-\mu_{1,n}}2
 \right\}.
\]
By the definition of $\varepsilon_n$, the conditional
mean separation has mean $\mu_{0,n}-\mu_{1,n}$ and
variance bounded as follows:
\begin{align*}
 &\Var_{D_n'}\!\left(
 \E_{P_{0,n}}[\hat\ell_n(X)\mid D_n']
 -
 \E_{P_{1,n}}[\hat\ell_n(X)\mid D_n']
 \right)\\
 &\quad\le
 2\left\{
 \Var_{D_n'}\!\left(
 \E_{P_{0,n}}[\hat\ell_n(X)\mid D_n']
 \right)
 +
 \Var_{D_n'}\!\left(
 \E_{P_{1,n}}[\hat\ell_n(X)\mid D_n']
 \right)
 \right\}\\
 &\quad=
 2\varepsilon_n(\mu_{0,n}-\mu_{1,n})^2.
\end{align*}
Thus, Chebyshev's inequality followed by a union bound gives
\begin{equation}\label{eq:reference-good}
 \P\bigl(
 H_n^c\cup\{|\hat\xi_{n/2}-\xi_n'|>u\}
 \bigr)
 \le
 8\varepsilon_n+\P(|\hat\xi_{n/2}-\xi_n'|>u).
\end{equation}
For the next steps, we condition on $(D_n',D_{n,1})$
and restrict to
$H_n\cap\{|\hat\xi_{n/2}-\xi_n'|\le u\}$.

\paragraph{Step 2: Negative drift on both sides of the changepoint.}
For $t\ge\xi_n'+R$, since $R>u$, we have
$t>\hat\xi_{n/2}$. The block
$\{\hat\xi_{n/2}+1,\ldots,t\}$ contains
$(\xi_n'-\hat\xi_{n/2})_+$ pre-change observations. Hence,
\begin{align*}
 &\E\left[
 \frac1{t-\hat\xi_{n/2}}
 \sum_{i=\hat\xi_{n/2}+1}^{t}\hat\ell_n(Y_{i,n})
 \,\middle|\,D_n',D_{n,1}
 \right]\\
 &\quad=
 \E_{P_{1,n}}[\hat\ell_n(X)\mid D_n']
 +
 \frac{(\xi_n'-\hat\xi_{n/2})_+}{t-\hat\xi_{n/2}}
 \left\{
 \E_{P_{0,n}}[\hat\ell_n(X)\mid D_n']
 -
 \E_{P_{1,n}}[\hat\ell_n(X)\mid D_n']
 \right\}.
\end{align*}
Similarly,
\begin{align*}
 &\E[\hat\mu_{t,n,L}\mid D_n',D_{n,1}]=
 \E_{P_{1,n}}[\hat\ell_n(X)\mid D_n']
 +
 \frac{\xi_n'}t
 \left\{
 \E_{P_{0,n}}[\hat\ell_n(X)\mid D_n']
 -
 \E_{P_{1,n}}[\hat\ell_n(X)\mid D_n']
 \right\}.
\end{align*}
Since $t>\hat\xi_{n/2}$, we have
\[
 \Delta_{t,\hat\xi_{n/2}}
 =\frac{S_n(\bY_n,t)}{|t-\hat\xi_{n/2}|}
 -\hat\mu_{t,n,L}= \frac1{t-\hat\xi_{n/2}}
\sum_{i=\hat\xi_{n/2}+1}^{t}\hat\ell_n(Y_{i,n})-\hat\mu_{t,n,L}.
\]
We have $\xi_n'/t\ge\xi_n'/m\ge c$ and, on the
training event under consideration,
\[
 \frac{(\xi_n'-\hat\xi_{n/2})_+}{t-\hat\xi_{n/2}}
 \le\frac{u}{R-u}\le\frac c2,
\]
where the last inequality uses~\eqref{eq:R-u}.
Together, these give
\begin{align}
 &\E[\Delta_{t,\hat\xi_{n/2}}\mid D_n',D_{n,1}]
 \nonumber\\
 &\quad=
 -
 \left\{
 \frac{\xi_n'}t
 -
 \frac{(\xi_n'-\hat\xi_{n/2})_+}{t-\hat\xi_{n/2}}
 \right\}
 \left\{
 \E_{P_{0,n}}[\hat\ell_n(X)\mid D_n']
 -
 \E_{P_{1,n}}[\hat\ell_n(X)\mid D_n']
 \right\}
 \nonumber\\
 &\quad\le
 -\frac c4(\mu_{0,n}-\mu_{1,n}).
 \label{eq:drift-right}
\end{align}

An analogous calculation for $t\le\xi_n'-R$ gives
\begin{align}
 &\E[\Delta_{t,\hat\xi_{n/2}}\mid D_n',D_{n,1}]
 \nonumber\\
 &\quad=
 -
 \left\{
 \frac{m-\xi_n'}{m-t}
 -
 \frac{(\hat\xi_{n/2}-\xi_n')_+}{\hat\xi_{n/2}-t}
 \right\}
 \left\{
 \E_{P_{0,n}}[\hat\ell_n(X)\mid D_n']
 -
 \E_{P_{1,n}}[\hat\ell_n(X)\mid D_n']
 \right\}
 \nonumber\\
 &\quad\le
 -\frac c4(\mu_{0,n}-\mu_{1,n}).
 \label{eq:drift-left}
\end{align}
Thus, conditional on $(D_n',D_{n,1})$,
$\Delta_{t,\hat\xi_{n/2}}$ has negative mean on either side.
Next, we show that the corresponding bound also holds
with high probability.

\paragraph{Step 3: Uniform control of fluctuations around the drift.}
For $i\in[m]$, define
$W_{i,n}=\hat\ell_n(Y_{i,n})
-\E[\hat\ell_n(Y_{i,n})\mid D_n',D_{n,1}]$.
For $t\ge\xi_n'+R$,
\[
 \Delta_{t,\hat\xi_{n/2}}
 -
 \E[\Delta_{t,\hat\xi_{n/2}}\mid D_n',D_{n,1}]
 =
 \frac{\sum_{i=\hat\xi_{n/2}+1}^{t}W_{i,n}}
      {t-\hat\xi_{n/2}}
 -
 \frac{\sum_{i=1}^{t}W_{i,n}}t.
\]
The two averages contain at least $R-u$ and
$\xi_n'\ge cm$ summands, respectively. These summands
are conditionally independent but need not be identically
distributed. Therefore, the general independent-variable
bound in Lemma~\ref{lem:maximal} gives
\begin{align*}
 &\P\left(
 \sup_{\substack{t\in[m-1]\\t\ge\xi_n'+R}}
 \left|
 \frac{\sum_{i=\hat\xi_{n/2}+1}^{t}W_{i,n}}
      {t-\hat\xi_{n/2}}
 \right|
 >
 \frac{c(\mu_{0,n}-\mu_{1,n})}{16}
 \,\middle|\,D_n',D_{n,1}
 \right)\le
 \frac{
 2^{10}\max_{j=0,1}
 \Var_{P_{j,n}}(\hat\ell_n(X)\mid D_n')}
 {c^2(\mu_{0,n}-\mu_{1,n})^2(R-u)}.
\end{align*}
Similarly, we derive
\begin{align*}
 &\P\left(
 \sup_{\substack{t\in[m-1]\\t\ge\xi_n'+R}}
 \left|
 \frac{\sum_{i=1}^{t}W_{i,n}}t
 \right|
 >
 \frac{c(\mu_{0,n}-\mu_{1,n})}{16}
 \,\middle|\,D_n',D_{n,1}
 \right)\le
 \frac{
 2^{10}\max_{j=0,1}
 \Var_{P_{j,n}}(\hat\ell_n(X)\mid D_n')}
 {c^2(\mu_{0,n}-\mu_{1,n})^2cm}.
\end{align*}
Together, these imply
\begin{align*}
 &\P\left(
 \sup_{\substack{t\in[m-1]\\t\ge\xi_n'+R}}
 \left|
 \Delta_{t,\hat\xi_{n/2}}
 -
 \E[\Delta_{t,\hat\xi_{n/2}}\mid D_n',D_{n,1}]
 \right|
 >
 \frac{c(\mu_{0,n}-\mu_{1,n})}{8}
 \,\middle|\,D_n',D_{n,1}
 \right)\\
 &\quad\le
 \frac{
 2^{10}\max_{j=0,1}
 \Var_{P_{j,n}}(\hat\ell_n(X)\mid D_n')}
 {c^2(\mu_{0,n}-\mu_{1,n})^2}
 \left(\frac1{R-u}+\frac1{cm}\right).
\end{align*}
For $t\le\xi_n'-R$, the same argument, applied to
the reversed sequences, gives
\begin{align*}
 &\P\biggl(\,\,
 \sup_{\substack{t\in[m-1]\\t\le\xi_n'-R}}
 \left|
 \Delta_{t,\hat\xi_{n/2}}
 -
 \E[\Delta_{t,\hat\xi_{n/2}}\mid D_n',D_{n,1}]
 \right|
 >
 \frac{c(\mu_{0,n}-\mu_{1,n})}{8}
 \,\biggm|\,D_n',D_{n,1}
 \biggr)\\
 &\quad\le
 \frac{
 2^{10}\max_{j=0,1}
 \Var_{P_{j,n}}(\hat\ell_n(X)\mid D_n')}
 {c^2(\mu_{0,n}-\mu_{1,n})^2}
 \left(\frac1{R-u}+\frac1{cm}\right).
\end{align*}
By the law of total variance,
\[
 \E\left[
 \max_{j=0,1}
 \Var_{P_{j,n}}(\hat\ell_n(X)\mid D_n')
 \right]
 \le\sigma_{0,n}^2+\sigma_{1,n}^2.
\]
Combining these bounds with
\eqref{eq:drift-right}--\eqref{eq:drift-left}, we obtain
\begin{align}
 &\P\biggl(\,\,
 \sup_{\substack{t\in[m-1]\\|t-\xi_n'|\ge R\\
                 t\ne\hat\xi_{n/2}}}
 \Delta_{t,\hat\xi_{n/2}}
 >
 -\frac{c(\mu_{0,n}-\mu_{1,n})}{8},\,
 H_n,\,
 |\hat\xi_{n/2}-\xi_n'|\le u
 \biggr)\le
 \frac{2^{11}(\sigma_{0,n}^2+\sigma_{1,n}^2)}
 {c^2(\mu_{0,n}-\mu_{1,n})^2}
 \left(\frac1{R-u}+\frac1{cm}\right).
 \label{eq:uniform-drift}
\end{align}

\paragraph{Step 4: Bound all remaining permutation $p$-values.}
Define
\[
 \mathcal G_n=
 \Biggl\{\,\,
 \sup_{\substack{t\in[m-1]\\|t-\xi_n'|\ge R\\
                 t\ne\hat\xi_{n/2}}}
 \Delta_{t,\hat\xi_{n/2}}
 >
 -\frac{c(\mu_{0,n}-\mu_{1,n})}{8}
 \Biggr\}.
\]
On
$H_n\cap\{|\hat\xi_{n/2}-\xi_n'|\le u\}\cap\mathcal G_n^c$,
every $t\in[m-1]$ with $|t-\xi_n'|\ge R$ satisfies
\[
 |t-\hat\xi_{n/2}|\ge R-u,
 \qquad
 \Delta_{t,\hat\xi_{n/2}}
 \le-\frac{c(\mu_{0,n}-\mu_{1,n})}{8}.
\]
For $t\ge\xi_n'+R$, we have $m/t\le1/c$;
for $t\le\xi_n'-R$, we have $m/(m-t)\le1/c$.
Lemma~\ref{lem:permutation} therefore gives, on both sides,
\[
 V^\pi_{t,n}\le\frac{\hat v_n}{c(R-u)},
 \qquad
 p_{t,n}
 \le
 \frac{\hat v_n/\{c(R-u)\}}
 {\hat v_n/\{c(R-u)\}
  +c^2(\mu_{0,n}-\mu_{1,n})^2/64}.
\]
Consequently, $p_{t,n}>\alpha$ implies
\[
 \hat v_n>
 \frac{
 \alpha c^3(R-u)(\mu_{0,n}-\mu_{1,n})^2}
 {64(1-\alpha)}.
\]
By Markov's inequality, the probability of this event
is at most
\[
 \frac{64(1-\alpha)}{\alpha c^3(R-u)}
 \left\{
 \frac{\sigma_{0,n}^2+\sigma_{1,n}^2}
      {(\mu_{0,n}-\mu_{1,n})^2}
 +\frac14
 \right\}.
\]
Combining this bound with~\eqref{eq:reference-good}
and~\eqref{eq:uniform-drift}, we obtain
\begin{align}
 \P\left(
 \max_{\substack{t\in[m-1]\\|t-\xi_n'|\ge R}}
 p_{t,n}>\alpha
 \right)&\le
 8\varepsilon_n+\P(|\hat\xi_{n/2}-\xi_n'|>u)
 \nonumber\\
 &+
 \frac{2^{11}(\sigma_{0,n}^2+\sigma_{1,n}^2)}
 {c^2(\mu_{0,n}-\mu_{1,n})^2}
 \left(\frac1{R-u}+\frac1{cm}\right)
 +
 \frac{64(1-\alpha)}{\alpha c^3(R-u)}
 \left\{
 \frac{\sigma_{0,n}^2+\sigma_{1,n}^2}
      {(\mu_{0,n}-\mu_{1,n})^2}
 +\frac14
 \right\}.
 \label{eq:finite-bound}
\end{align}

\paragraph{Step 5: Choose the exclusion radius.}
Recall $R=\lceil Mr_n\rceil$ and
$u=\lfloor cR/(2+c)\rfloor$. For sufficiently large
$M$, $u\ge2$, and both $u-1$ and $R-u$ are bounded
below by positive multiples of $Mr_n$, depending only
on $c$. By~\eqref{eq:SNR},
\[
 \frac{\sigma_{0,n}^2+\sigma_{1,n}^2}
      {(\mu_{0,n}-\mu_{1,n})^2}
 \le r_n.
\]
For the reference estimator in~\eqref{eq:MLE},
Proposition~\ref{prop:MLE} now bounds~\eqref{eq:finite-bound}
by
\[
 \left(8+\frac4{\kappa^2}\right)\varepsilon_n
 +
 C_{c,\kappa}
 \left(\frac1{\alpha M}+\frac{r_n}{m}\right),
\]
for a finite constant $C_{c,\kappa}$ depending only on
$c$ and $\kappa$. By
Assumption~\ref{ass:learned-score}(iii) and $r_n/n\to0$,
first let $n\to\infty$ and then $M\to\infty$
to obtain~\eqref{eq:sharpness}.

On the event that all candidates at distance at least
$R$ are excluded, the interval hull of the retained calibration locations
contains at most $2\lceil Mr_n\rceil+1$ locations. Therefore,
$|\bar\Ccal^{\conch}_{n,1-\alpha}|
\le4\lceil Mr_n\rceil+2$.
This proves $|\bar\Ccal^{\conch}_{n,1-\alpha}|=\Op(r_n)$.
Since $r_n/n\to0$, its normalized cardinality converges
to zero.
\hfill$\square$

\subsubsection{Auxiliary lemma}

\begin{lemma}\label{lem:permutation}
Conditional on $(D_n',D_{n,1},\bY_n)$, let $\pi$ be
uniform on $\Pi_t$ and write
\[
 V^\pi_{t,n}
 =
 \Var_\pi\left(
 \frac{S_n(\pi(\bY_n),t)}{|t-\hat\xi_{n/2}|}
 \,\middle|\,D_n',D_{n,1},\bY_n
 \right),
 \qquad t\ne\hat\xi_{n/2}.
\]
Then
\[
 \E_\pi\left[
 \frac{S_n(\pi(\bY_n),t)}{|t-\hat\xi_{n/2}|}
 \,\middle|\,D_n',D_{n,1},\bY_n
 \right]
 =
 \begin{cases}
 \hat\mu_{t,n,L},&t>\hat\xi_{n/2},\\
 \hat\mu_{t,n,R},&t<\hat\xi_{n/2},
 \end{cases}
\]
and the corresponding variance satisfies
\[
 V^\pi_{t,n}
 \le
 \begin{cases}
 \displaystyle
 \frac{m\hat v_n}{t(t-\hat\xi_{n/2})},
 &t>\hat\xi_{n/2},\\[1ex]
 \displaystyle
 \frac{m\hat v_n}{(m-t)(\hat\xi_{n/2}-t)},
 &t<\hat\xi_{n/2}.
 \end{cases}
\]
If $\Delta_{t,\hat\xi_{n/2}}<0$, then
$p_{t,n}\le
V^\pi_{t,n}/(V^\pi_{t,n}+\Delta_{t,\hat\xi_{n/2}}^2)$.
Further,
$\E[\hat v_n]\le
\sigma_{0,n}^2+\sigma_{1,n}^2
+(\mu_{0,n}-\mu_{1,n})^2/4$.
\end{lemma}

\begin{proof}
Suppose $t>\hat\xi_{n/2}$. By~\eqref{eq:score},
\[
 \frac{S_n(\pi(\bY_n),t)}{t-\hat\xi_{n/2}}
 =
 \frac1{t-\hat\xi_{n/2}}
 \sum_{i=\hat\xi_{n/2}+1}^{t}\hat\ell_n(Y_{\pi(i),n}).
\]
The set $\{\pi(\hat\xi_{n/2}+1),\ldots,\pi(t)\}$ is sampled
uniformly without replacement from $[t]$ and has size
$t-\hat\xi_{n/2}$. Since the expression above averages the
scores over these indices,
\begin{align}
 \E_\pi\left[
 \frac{S_n(\pi(\bY_n),t)}{t-\hat\xi_{n/2}}
 \,\middle|\,D_n',D_{n,1},\bY_n
 \right]
 &=\hat\mu_{t,n,L},
 \nonumber\\
 V^\pi_{t,n}
 &=
 \frac{\hat\xi_{n/2}}{(t-\hat\xi_{n/2})(t-1)}
 \left\{
 \frac1t\sum_{i=1}^{t}
 \bigl(
 \hat\ell_n(Y_{i,n})-\hat\mu_{t,n,L}
 \bigr)^2
 \right\}.
 \label{eq:variance-right}
\end{align}
Furthermore,
\[
 \frac1t\sum_{i=1}^{t}
 \bigl(
 \hat\ell_n(Y_{i,n})-\hat\mu_{t,n,L}
 \bigr)^2
 \le
 \frac1t\sum_{i=1}^{t}
 \left(
 \hat\ell_n(Y_{i,n})
 -
 \frac{\mu_{0,n}+\mu_{1,n}}2
 \right)^2
 \le\frac mt\hat v_n.
\]
Since $\hat\xi_{n/2}/(t-1)\le1$, combining this
with~\eqref{eq:variance-right} proves the stated mean
and variance bound when $t>\hat\xi_{n/2}$. The case
$t<\hat\xi_{n/2}$ follows analogously by considering the
negative average of $\hat\xi_{n/2}-t$ scores sampled
without replacement from $\{t+1,\ldots,m\}$.

By the definition of the permutation $p$-value,
\begin{multline*}
 p_{t,n}
 =
 \P_\pi\Biggl(
 \frac{S_n(\pi(\bY_n),t)}{|t-\hat\xi_{n/2}|}
 -
 \E_\pi\left[
 \frac{S_n(\pi(\bY_n),t)}{|t-\hat\xi_{n/2}|}
 \,\middle|\,D_n',D_{n,1},\bY_n
 \right]
 \le\Delta_{t,\hat\xi_{n/2}}
 \,\Biggm|\,D_n',D_{n,1},\bY_n
 \Biggr).
\end{multline*}
Cantelli's inequality therefore gives the bound on
$p_{t,n}$ whenever $\Delta_{t,\hat\xi_{n/2}}<0$.

Finally, for $i\le\xi_n'$, the unconditional mean of
$\hat\ell_n(Y_{i,n})$ is $\mu_{0,n}$, while for
$i>\xi_n'$ it is $\mu_{1,n}$. In either case, its
distance from $(\mu_{0,n}+\mu_{1,n})/2$ is
$(\mu_{0,n}-\mu_{1,n})/2$. Therefore,
\begin{align*}
 \E[\hat v_n]=
 \frac{\xi_n'}m
 \left\{
 \sigma_{0,n}^2
 +\frac{(\mu_{0,n}-\mu_{1,n})^2}{4}
 \right\}+
 \frac{m-\xi_n'}m
 \left\{
 \sigma_{1,n}^2
 +\frac{(\mu_{0,n}-\mu_{1,n})^2}{4}
 \right\},
\end{align*}
which proves the final part.
\end{proof}
\subsection{Proof of Corollary~\ref{cor:oracle}: oracle localization rates}

If $\hat\ell_n=\ell_n$, we observe that
\[
 \mu_{0,n}=\KL(P_{0,n}\Vert P_{1,n}),
 \qquad
 \mu_{1,n}=-\KL(P_{1,n}\Vert P_{0,n}),
 \qquad
 \varepsilon_n=0.
\]
Hence, by definition, $\mu_{0,n}-\mu_{1,n}=J_n$, and the assumptions
of the corollary imply
Assumption~\ref{ass:learned-score}.
Moreover,
\[
 r_n
 =
 1\vee
 \frac{\sigma_{0,n}^2+\sigma_{1,n}^2}{J_n^2}
 \le
 1\vee\frac{2C_v}{J_n}
 =
 O(J_n^{-1}).
\]
Since $nJ_n\to\infty$, we have $r_n/n\to0$.
Proposition~\ref{prop:MLE} and
Theorems~\ref{thm:sharpness} and~\ref{thm:clone-sharp}
give the stated oracle localization rates.
\hfill$\square$

\subsection{Proof of Proposition~\ref{prop:lower-fixed}: lower bound on cardinality}
We start by writing $d=(1-\alpha)^2/2$, and considering the interval $\mathcal J_n=[\xi_n-d/J_n,\xi_n+d/J_n]\cap\mathbb Z$. Since $J_n\to0$, $nJ_n\to\infty$, and $\xi_n/n\to\tau\in(0,1)$, for large $n$, it holds that $\mathcal J_n\subseteq\mathcal{T}$ and $|\mathcal J_n|\ge d/J_n$.

Recall that $\P_{t,n}=P_{0,n}^{\otimes t}\otimes P_{1,n}^{\otimes(n-t)}$. Moreover, KL between product measures is sum of KL computed between individual measures. For $t\in\mathcal J_n$, therefore we have that
\[
 \KL(\P_{\xi_n,n}\Vert\P_{t,n})
 \le |t-\xi_n|J_n\le d.
\]
Pinsker's inequality and~\eqref{eq:honesty-fixed} therefore give, for each $t\in\mathcal J_n$,
\[
 \P_{\xi_n,n}(t\in\widetilde\Ccal_n)
 \ge \P_{t,n}(t\in\widetilde\Ccal_n)
       -\TV(\P_{\xi_n,n},\P_{t,n})
 \ge (1-\alpha)-\sqrt{d/2}=(1-\alpha)/2.
\]
Thus, each $t\in\mathcal J_n$ can be missed by $\widetilde\Ccal_n$ with
probability at most $(1+\alpha)/2$. Summing over $t\in\mathcal J_n$ gives
\[
 \E_{\bX\sim \P_{\xi_n,n}}\bigl[|\mathcal J_n\setminus\widetilde\Ccal_n|\bigr]
 \le \frac{1+\alpha}{2}|\mathcal J_n|.
\]
Now, if $\widetilde\Ccal_n$ contains fewer than a fraction $(1-\alpha)/4$
of the locations in $\mathcal J_n$, then $\mathcal J_n\setminus\widetilde\Ccal_n$ must contain more than a fraction
$(3+\alpha)/4$ of locations in $\mathcal{J}_n$. Hence, by Markov's inequality,
\[
 \P_{\xi_n,n}\left(
   |\widetilde\Ccal_n\cap\mathcal J_n|
   < \frac{1-\alpha}{4}|\mathcal J_n|
 \right)
 \le \frac{(1+\alpha)/2}{(3+\alpha)/4}
 = \frac{2(1+\alpha)}{3+\alpha}.
\]
Consequently, with probability at least $(1-\alpha)/(3+\alpha)$,
\[
 |\widetilde\Ccal_n|
 \ge \frac{1-\alpha}{4}|\mathcal J_n|
 \ge \frac{(1-\alpha)d}{4J_n}
 = \frac{(1-\alpha)^3}{8J_n}.
\]
This proves~\eqref{eq:lower-fixed} with $c_\alpha=(1-\alpha)^3/8$
and $p_\alpha=(1-\alpha)/(3+\alpha)$.
\hfill$\square$

\section{Proof of results from Appendix~\ref{sec:selective_inference}}
\subsection{Proof of Proposition~\ref{prop:detection-valid}}
Under $\Hcal_{0,n}$, $\pi(\bX)\overset{d}{=}\bX$ for every $\pi\in\mathcal S_n$. Composition within $\mathcal S_n$ shows that the values $p_n(\pi(\bx))$ are normalized weak lower ranks of the scores $S(\pi(\bx),n)$. Hence, for every $u\in(0,1)$,
\[
 \P_{\Hcal_{0,n}}(p_n\le u)
 =\E_{\Hcal_{0,n}}\left[
 \frac1{|\mathcal S_n|}\sum_{\pi\in\mathcal S_n}
 \One{p_n(\pi(\bX))\le u}\right]\le u.
\]
Taking $u=\beta$ gives the detection bound.
\hfill$\square$

\subsection{Proof of Theorem~\ref{thm:selective}: finite-sample validity of \detectconch{}}\label{app:selection}

Suppose Assumption~\ref{ass:exchangeability} holds for $\bX$ with changepoint at $\xi\in[n-1]$. We treat the detection rule $A$ and the CPP score $S$ fixed.

For a fixed sequence $\bx$, write $M(\bx)=\sum_{\pi\in\Pi_\xi}\One{A(\pi(\bx))=1}$. If $A(\bx)=1$, then $M(\bx)\ge1$ since the identity permutation belongs to $\Pi_\xi$. For notational convenience, we set $p_\xi^{\mathrm{det}}(\bx)=1$ when $A(\bx)=0$. Note that this does not affect the procedure.

Now, fix any $\bx$ with $M(\bx)>0$. Then, for every $\pi\in\Pi_\xi$ such that $A(\pi(\bx))=1$, we can write
\[
p_\xi^{\mathrm{det}}(\pi(\bx))
=
\frac{
 \displaystyle\sum_{\sigma\in\Pi_\xi}
 \One{A(\sigma(\bx))=1}
 \One{S(\sigma(\bx),\xi)\le S(\pi(\bx),\xi)}
}{
 M(\bx)
}.
\]
In words, $p_\xi^{\mathrm{det}}(\pi(\bx))$ is the normalized rank of $S(\pi(\bx),\xi)$ among the collection $\{S(\sigma(\bx),\xi):\sigma\in\Pi_\xi, A(\sigma(\bx))=1\}$. As a result, by \cite[Lemma~3]{harrison2012conservative}, we derive 
\begin{align}\label{eq:valid-normalized-selective-rank}
\frac1{M(\bx)}
\sum_{\pi\in\Pi_\xi}
\One{
 A(\pi(\bx))=1,\,
 p_\xi^{\mathrm{det}}(\pi(\bx))\le\alpha
}
&\le
\alpha\quad \text{and thus,}\nonumber\\
   \frac1{|\Pi_\xi|}
\sum_{\pi\in\Pi_\xi}
\One{
 A(\pi(\bx))=1,\,
 p_\xi^{\mathrm{det}}(\pi(\bx))\le\alpha
}
&\le
\frac{\alpha M(\bx)}{|\Pi_\xi|}.
\end{align}
Note that by construction, the same inequality holds when $M(\bx)=0$, since both sides
are then zero.

Finally, since $\bX\overset{d}{=}\pi(\bX)$ for every $\pi\in\Pi_\xi$, by~\eqref{eq:valid-normalized-selective-rank}, we obtain that
\begin{align*}
\P_\xi\bigl(
 A(\bX)=1,\,
 p_\xi^{\mathrm{det}}(\bX)\le\alpha
\bigr)
&=\frac1{|\Pi_\xi|}
\sum_{\pi\in\Pi_\xi}
\P_\xi\bigl(
 A(\pi(\bX))=1,\,
 p_\xi^{\mathrm{det}}(\pi(\bX))\le\alpha
\bigr)\\
&\quad=
\E_\xi\left[
 \frac1{|\Pi_\xi|}
 \sum_{\pi\in\Pi_\xi}
 \One{
  A(\pi(\bX))=1,\,
  p_\xi^{\mathrm{det}}(\pi(\bX))\le\alpha
 }
\right]\\
&\quad\le
\frac{\alpha}{|\Pi_\xi|}
\sum_{\pi\in\Pi_\xi}
\P_\xi\bigl(A(\pi(\bX))=1\bigr)
=
\alpha\,\P_\xi\bigl(A(\bX)=1\bigr).
\end{align*}

On the event $\{A(\bX)=1\}$ , the event $\xi\notin\Ccal^{\detectconch}_{n,1-\alpha}$ is exactly the event $p_\xi^{\mathrm{det}}(\bX)\le\alpha$. Hence, we get
\[
\P_\xi\bigl(
 A(\bX)=1,\,
 \xi\notin\Ccal^{\detectconch}_{n,1-\alpha}
\bigr)
\le
\alpha\,\P_\xi\bigl(A(\bX)=1\bigr).
\]
Whenever the detection probability is positive, it further yields
\[
\P_\xi\bigl(
 \xi\in\Ccal^{\detectconch}_{n,1-\alpha}
 \mid A(\bX)=1
\bigr)\ge1-\alpha.
\]
\hfill$\square$

\end{document}